\documentclass[%
  reprint,
  superscriptaddress,
  amsmath,amssymb,
  aps,
  prb,
]{revtex4-1}

\usepackage{silence}
\allowdisplaybreaks

\usepackage{graphicx}
\usepackage{url}
\usepackage[dvipsnames]{xcolor}
\usepackage{tikz}
\usetikzlibrary{
  decorations.markings,
  decorations.pathmorphing,
  arrows.meta
  }

\usepackage{hyperref}
\hypersetup{
	setpagesize=false,
	bookmarksnumbered=true,%
	bookmarksopen=true,%
	colorlinks=true,%
	linkcolor=magenta,
	citecolor=magenta,
}

\usepackage{bbm,mathrsfs}
\usepackage{makecell}
\usepackage{dcolumn}
\usepackage{multirow}
\usepackage{ulem}
\usepackage{booktabs}
\AtBeginDocument{%
  \heavyrulewidth=.08em
  \lightrulewidth=.05em
  \cmidrulewidth=.03em
  \belowrulesep=.65ex
  \belowbottomsep=0pt
  \aboverulesep=.4ex
  \abovetopsep=0pt
  \cmidrulesep=\doublerulesep
  \cmidrulekern=.5em
  \defaultaddspace=.5em
}
\usepackage{braket}
\usepackage[capitalise]{cleveref}
\crefname{section}{Sec.}{Section}
\Crefname{section}{Section}{Sections}
\crefname{table}{Tab.}{Tab.}

\usepackage[version=4]{mhchem}
\usepackage{physics}

\usepackage[whole]{bxcjkjatype}
\usepackage{natbib}

\input{unicode-math-pdflatex.sty}

\newcommand{\DD}[1]{\mathcal{D}{#1}}
\newcommand{\QPT}{\mathrm{QPT}}
\newcommand{\ET}{\mathrm{ET}}

\DeclareMathOperator{\sgn}{sgn}
\DeclareMathOperator{\Pf}{Pf}
\DeclareMathOperator{\qhc}{h.c.}

\begin{document} 
\preprint{APS/123-QED}

\title{
  Phase-shift instanton approach to tunneling duality in Read--Rezayi state
}
\author{Ryoi Ohashi}
\affiliation{
  Department of Physics, Kyushu University, Fukuoka 819-0395, Japan
}
\author{Hiroki Isobe}
\affiliation{
  Department of Physics, Kyushu University, Fukuoka 819-0395, Japan
}
\author{Ryota Nakai}
\affiliation{
  RIKEN Center for Quantum Computing (RQC), Wako, Saitama 351-0198, Japan
}
\author{Kentaro Nomura}
\affiliation{
  Department of Physics, Kyushu University, Fukuoka 819-0395, Japan
}
\affiliation{
  Quantum and Spacetime Research Institute, Kyushu University, Fukuoka 819-0395, Japan
}
\date{\today}

\begin{abstract}
  We study the duality between quasi-particle and electron tunneling in point-contact geometries of fractional quantum Hall states.
  To treat non-Abelian edge operators, we introduce a “phase-shift instanton” that incorporates phase factors from primary fields into the instanton gas framework.
  Using this method, we reformulate the Moore--Read duality and obtain an explicit dual description for the $k=3$ Read-Rezayi state.
  Our results clarify how quasi-particle tunneling produces characteristic phase shifts in instantons and how these shifts map strong quasi-particle tunneling to weak electron tunneling.
  Based on this dual description, we analytically evaluate the non-linear differential conductance in the strong-coupling regime.
  We reveal that, due to the physical requirement that the tunneling particle across the vacuum gap must be a true fermion, the transport behavior universally converges to a $G \propto V^4$ scaling for both the Moore--Read and Read--Rezayi states.
  This universal transport signature highlights a fundamental topological constraint underlying non-Abelian fractional quantum Hall edges.
\end{abstract}
\maketitle

\section{Introduction}
\label{sec:introduction}

The fractional quantum Hall (FQH) effect in two-dimensional electron systems under strong magnetic fields produces incompressible quantum fluids whose excitations carry fractional charge.\cite{klitzing_1980_New,tsui_1982_TwoDimensional,willett_1987_Observation,prange_1990_Quantum} 
These quasi-holes are known to behave as anyons, which can be Abelian, acquiring a phase under exchange, or non-Abelian, where exchanges act noncommutatively on internal fusion degrees of freedom.
The Laughlin wave functions describe the simplest Abelian FQH states featuring fractionally charged anyons.\cite{laughlin_1983_Anomalous,haldane_1985_Periodic,jain_1989_Compositefermion,jain_1989_Incompressible,jain_1990_Theory} 
The Moore--Read and Read--Rezayi states exhibit non-Abelian pairing and clustering patterns, respectively,\cite{moore_1991_Nonabelions,read_1999_Paired} and have attracted significant interest for potential applications in topological quantum computation.\cite{kitaev_2003_Faulttolerant,freedman_2002_Modular,nayak_2008_NonAbelian,mong_2014_Universal}

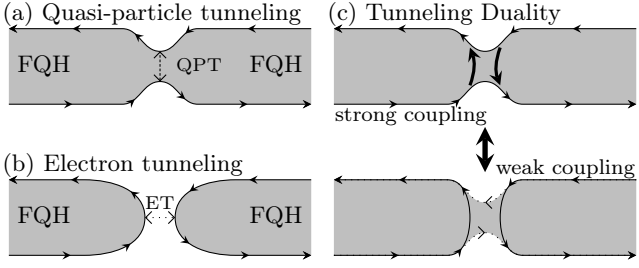
\begin{figure}[htbp]
  \centering
  \begin{tikzpicture}
    \begin{scope}[shift={(0,2)}]
      \node [anchor=west] (note) at (-0.2,1.2) {(a) Quasi-particle tunneling};
      \fill[lightgray](0,0) -- (1.5, 0) to[out=0, in=180] (2,0.3) to [out=0, in=180] (2.5,0) -- (4,0) -- (4,1) -- (2.5,1) to[out=180, in=0] (2,0.7) to[out=180, in=0] (1.5,1) -- (0,1) -- cycle;
      \draw[-stealth,postaction=decorate,
        decoration={markings,
        mark = between positions 0.2 and 1 step 0.2
        with {\arrow{stealth}}}]
        (0, 0) -- (1.5, 0) to[out=0, in=180] (2, 0.3) to[out=0, in=180] (2.5,0) -- (4, 0);
      \draw[-stealth,postaction=decorate,
        decoration={markings,
        mark = between positions 0.2 and 1 step 0.2
        with {\arrow{stealth}}}]
      (4, 1) -- (2.5, 1) to[out=180, in=0] (2, 0.7) to[out=180, in=0] (1.5,1) -- (0, 1);

      \node [anchor=west] (note) at (0,0.5) {FQH};
      \node [anchor=east] (note) at (4,0.5) {FQH};

      \draw[thin,dotted,dash pattern=on 1pt off 0.7pt,<->,>=Straight Barb] (2, 0.3) -- (2, 0.7);
      \node [anchor=west,font=\scriptsize] at (2.1,0.5) {QPT};
    \end{scope}

    \begin{scope}[shift={(0,0)}]
      \node [anchor=west] (note) at (-0.2,1.2) {(b) Electron tunneling};
      \fill[lightgray](0, 0) -- (1, 0) to[out=0, in=-90] (1.8, 0.5) to[out=90, in=0] (1,1) -- (0, 1) -- cycle;
      \fill[lightgray](3, 1) to [out=180, in=90] (2.2, 0.5) to[out=-90, in=180] (3,0) -- (4, 0) -- (4, 1) -- cycle;

      \draw[-stealth,postaction=decorate,
        decoration={markings,
        mark = between positions 0.2 and 1 step 0.2
        with {\arrow{stealth}}}]
      (0, 0) -- (1, 0) to[out=0, in=-90] (1.8, 0.5) to[out=90, in=0] (1,1) -- (0, 1);
      \draw[-stealth,postaction=decorate,
        decoration={markings,
        mark = between positions 0.2 and 1 step 0.2
        with {\arrow{stealth}}}]
      (4, 1) -- (3, 1) to[out=180, in=90] (2.2, 0.5) to[out=-90, in=180] (3,0) -- (4, 0);

      \node [anchor=west] (note) at (0,0.5) {FQH};
      \node [anchor=east] (note) at (4,0.5) {FQH};

      \draw[thin,dotted,<->,>=Straight Barb] (1.8, 0.5) -- (2.2, 0.5);
      \node [anchor=south,font=\scriptsize] at (2,0.5) {ET};
    \end{scope}

    \begin{scope}[shift={(4.3,2)}]
      \node [anchor=west] (note) at (-0.2,1.2) {(c) Tunneling Duality};
      \begin{scope}[shift={(0,0)}]
        \fill[lightgray](0,0) -- (1.5, 0) to[out=0, in=180] (2,0.3) to [out=0, in=180] (2.5,0) -- (4,0) -- (4,1) -- (2.5,1) to[out=180, in=0] (2,0.7) to[out=180, in=0] (1.5,1) -- (0,1) -- cycle;
        \draw[-stealth,postaction=decorate,
          decoration={markings,
          mark = between positions 0.2 and 1 step 0.2
          with {\arrow{stealth}}}]
        (0, 0) -- (1.5, 0) to[out=0, in=180] (2, 0.3) to[out=0, in=180] (2.5,0) -- (4, 0);
        \draw[-stealth,postaction=decorate,
          decoration={markings,
          mark = between positions 0.2 and 1 step 0.2
          with {\arrow{stealth}}}]
        (4, 1) -- (2.5, 1) to[out=180, in=0] (2, 0.7) to[out=180, in=0] (1.5,1) -- (0, 1);

        \draw[line width=1.2pt,<-,>=stealth] (1.8,0.75) to[in=70,out=-70] (1.8,0.25); 
        \draw[line width=1.2pt,->,>=stealth] (2.2,0.75) to[in=110,out=-110] (2.2,0.25); 
      \end{scope}

      \begin{scope}[shift={(0,-0.6)}]
        \node [anchor=west,font=\footnotesize] (note) at (-0.1,0.45) {strong coupling};
        \draw[line width=2pt,<->,>=stealth] (2,0.3) -- (2.0,-0.3); 
        \node [anchor=east,font=\footnotesize] (note) at (4.1,-0.25) {weak coupling};
      \end{scope}

      \begin{scope}[shift={(0,-2)}]
        \fill[lightgray](0,0) -- (1.5, 0) to[out=0, in=180] (2,0.3) to [out=0, in=180] (2.5,0) -- (4,0) -- (4,1) -- (2.5,1) to[out=180, in=0] (2,0.7) to[out=180, in=0] (1.5,1) -- (0,1) -- cycle;
        \draw[dotted,postaction=decorate,
          decoration={markings,
          mark = at position 0.5
          with \arrow{Straight Barb}}]
        (0, 0) -- (1.5, 0) to[out=0, in=180] (2, 0.3) to[out=0, in=180] (2.5,0) -- (4, 0);
        \draw[dotted,postaction=decorate,
          decoration={markings,
          mark = at position 0.5
          with \arrow{Straight Barb}}]
        (4, 1) -- (2.5, 1) to[out=180, in=0] (2, 0.7) to[out=180, in=0] (1.5,1) -- (0, 1);

        \draw[-stealth,postaction=decorate,
          decoration={markings,
          mark = between positions 0.2 and 1 step 0.2
          with {\arrow{stealth}}}]
        (0, 0) -- (1.5, 0) to[out=0, in=-90] (1.8, 0.5) to[out=90, in=0] (1.5,1) -- (0, 1);
        \draw[-stealth,postaction=decorate,
          decoration={markings,
          mark = between positions 0.2 and 1 step 0.2
          with {\arrow{stealth}}}]
        (4, 1) -- (2.5, 1) to[out=180, in=90] (2.2, 0.5) to[out=-90, in=180] (2.5,0) -- (4, 0);

      \end{scope}
    \end{scope}
  \end{tikzpicture}
  \caption{
    Illustration of the point-contact geometry in FQH systems. 
    (a) and (b) show QPT and ET processes at a point contact, respectively. 
    (c) highlights the tunneling duality. 
  }
  \label{fig: MR_Geometry}
\end{figure}

A central theoretical tool for understanding these phases is the bulk-edge correspondence: bulk topological orders manifest in chiral boundary theories whose operator content encodes the charge and statistics of bulk excitations.\cite{witten_1989_Quantum,witten_1992_Holomorphic,stone_1991_Vertex,wen_1991_EDGE,wen_1995_Topological,fradkin_2013_Field,chang_2003_Chiral}
Experimentally, point-contact geometries, as illustrated in \cref{fig: MR_Geometry}(a) and (b), which permit quasi-particle and electron tunneling between edges, have proven to be powerful probes of edge structure and fractional charge.\cite{milliken_1996_Indications,roddaro_2004_Interedge,roddaro_2004_Quasiparticle,picciotto_1997_Direct,saminadayar_1997_Observation,veillon_2024_Observation,miller_2007_Fractional,radu_2008_QuasiParticle,dolev_2008_Observation,bid_2010_Observation,venkatachalam_2011_Local}
Theoretical analysis of tunneling typically relies on perturbative methods, such as renormalization group (RG) arguments, and instanton techniques.\cite{schmid_1983_Diffusion,fisher_1985_Quantum,callan_1992_Phase,kane_1992_Transmission,kane_1992_Transport,fendley_1995_Exact,dec.chamon_1997_Distinct,imura_1997_Quantum,furusaki_1993_Resonant,nomura_2001_Strong,fendley_2006_Dynamical,fendley_2007_Edge,fendley_2009_Boundary,ito_2012_QuasiParticle,ahmed_2026_Universal}

An important concept in the theory of point-contact tunneling is duality, illustrated in \cref{fig: MR_Geometry}(c): strong-coupling quasi-particle tunneling (QPT) can be mapped onto a weak-coupling electron tunneling (ET) problem allowing nonperturbative regimes to be accessed via a dual weakly coupled description.
Once the problem can be mapped to a weak‑coupling description, subsequent evaluations can be performed within the perturbative regime, making this mapping a highly powerful technique for qualitatively analysing tunneling processes.
This duality is known for Abelian Laughlin states\cite{schmid_1983_Diffusion,fisher_1985_Quantum,callan_1992_Phase,kane_1992_Transmission,imura_1997_Quantum,furusaki_1993_Resonant} and non-Abelian Moore–Read states.\cite{nomura_2001_Strong,fendley_2006_Dynamical,fendley_2007_Edge,fendley_2009_Boundary,ito_2012_QuasiParticle,ahmed_2026_Universal}
However, extending this duality to more complex non-Abelian states, such as the Read--Rezayi states, remains an open challenge due to the intricate operator algebra of their edge theories.

In this work, we introduce the notion of a ``phase-shift instanton'' to incorporate phase factors arising from non-Abelian primary fields into the instanton gas formulation.
Using this construction, we reformulate the Moore--Read duality within a phase-shift instanton framework and present the first explicit application of this approach to the $k=3$ Read--Rezayi state.
While a fully rigorous mathematical proof of this duality transformation remains an open challenge, our approach provides a physically consistent framework.

Crucially, this dual description enables us to analytically evaluate the transport properties in the previously inaccessible strong-coupling regime.
Confirming its validity, our method successfully reproduces the known $G \propto V^4$ scaling for the Moore--Read state.
Furthermore, we reveal that the Read--Rezayi state also exhibits exactly the same $G \propto V^4$ scaling in the strong-coupling limit.
This striking universality emerges from the physical requirement that the single particle tunneling across the vacuum gap must be a true fermion, demonstrating how complex non-Abelian fractionalization reconstructs the universal electron under the duality mapping.

Our paper is organized as follows.
In \cref{sec:formulation_and_model} we summarize the bulk and edge models for the Laughlin, Moore--Read, and Read--Rezayi states and introduce the point-contact tunneling Hamiltonian.
\Cref{sec:duality_Laughlin} reviews the duality in Laughlin states via the instanton approach.
In \cref{sec:duality_MR} we introduce the phase-shift instanton for the Moore--Read state and reexamine its duality.
\Cref{sec:duality_RR} applies the same approach to the Read--Rezayi state.
In \cref{sec:conductance}, we discuss the differential conductance behavior based on this duality to connect with experiment.
Finally, we conclude in \cref{sec:conclusion}, with technical derivations collected in Appendices.

\section{Formulation and Model} 
\label{sec:formulation_and_model}
In this section, we introduce bulk and edge theories of three FQH states and construct the tunneling Hamiltonian in the point-contact geometry.
\begin{table}[htp]
  \centering
  \caption{
    Summary of the filling factors, fractional charges, and corresponding quasi-particle and electron operators for the Laughlin, Moore--Read, and $k=3$ Read--Rezayi states.
  }
  \label{tab:operators}
  \begin{tabular}{lccc}
    \toprule
    & Laughlin & Moore--Read & Read--Rezayi \\
    \midrule
    \midrule
    Filling factor 
      & $\phantom{{}^{\ast}}ν=\dfrac{1}{2m+1}$ 
      & $\phantom{{}^{\ast}}ν=\dfrac{1}{2}$ 
      & $\phantom{{}^{\ast}}ν=\dfrac{3}{5}$ \\
    \midrule
    Fractional  
      & \multirow{2}{*}{$e^{\ast}=\dfrac{e}{2m+1}$}
      & \multirow{2}{*}{$e^{\ast}=\dfrac{e}{4}$}
      & \multirow{2}{*}{$e^{\ast}=\dfrac{e}{5}$}\\
    \hspace{6mm}\phantom{oi}charge & & & \\
    \midrule
    Quasi-hole
      & \multirow{2}{*}{$Ψ_\mathrm{QH}^\mathrm{Lau}=e^{iϕ\phantom{/1}}$ }
      & \multirow{2}{*}{$Ψ_\mathrm{QH}^\mathrm{MR}=σ e^{iϕ/2}$ }
      & \multirow{2}{*}{$Ψ_\mathrm{QH}^\mathrm{RR}=σ₁ e^{iϕ/3}$} \\
    \hspace{6mm}operator & & & \\
    \midrule
    Electron 
      & \multirow{2}{*}{$Ψ_\mathrm{el}^\mathrm{Lau}=e^{iϕ/ν}$} 
      & \multirow{2}{*}{$Ψ_\mathrm{el}^\mathrm{MR}=ψ e^{iϕ/ν}$} 
      & \multirow{2}{*}{$Ψ_\mathrm{el}^\mathrm{RR}=ψ₁ e^{iϕ/ν}$}\\
    \hspace{6mm}operator & & & \\
    \bottomrule
  \end{tabular}
\end{table}

\begin{table}[tbp]
  \centering
  \caption{
    List of the quasi-particle and electron operator candidates cataloged by their fractional charge $Q$ for the Laughlin, Moore--Read, and k=3 Read--Rezayi states.
    }
  \label{tab:quasi-particle_operators}
  \begin{tabular}{lll}
    \toprule
    FQH state & Charge $Q$ & Operator(s) \\
    \midrule\midrule
    \multirow{4}{*}{Laughlin ($\nu=\frac{1}{2m+1}$)} 
    & $\frac{1}{2m+1}e$ & $e^{i\phi}$ \\
    & $\frac{2}{2m+1}e$ & $e^{i2\phi}$ \\
    & \multicolumn{1}{l}{\vdots} & \multicolumn{1}{l}{\vdots} \\
    & $e$ & $e^{i(2m+1)\phi}$ \\
    \midrule
    \multirow{4}{*}{Moore--Read ($\nu=1/2$)} 
    & $e/4$ & $\sigma e^{i\phi/2}$ \\
    & $e/2$ & $I e^{i\phi}, \psi e^{i\phi}$ \\
    & $3e/4$ & $\sigma e^{i3\phi/2}$ \\
    & $e$ & $I e^{i2\phi}, \psi e^{i2\phi}$ \\
    \midrule
    \multirow{5}{*}{Read--Rezayi ($\nu=3/5$)} 
    & $e/5$ & $\sigma_1 e^{i\phi/3}, \psi_2 e^{i\phi/3}$ \\
    & $2e/5$ & $\sigma_2 e^{i2\phi/3}, \psi_1 e^{i2\phi/3}$ \\
    & $3e/5$ & $I e^{i\phi}, \epsilon e^{i\phi}$ \\
    & $4e/5$ & $\sigma_1 e^{i4\phi/3}, \psi_2 e^{i4\phi/3}$ \\
    & $e$ & $\psi_1 e^{i5\phi/3}, \sigma_2 e^{i5\phi/3}$ \\
    \bottomrule
  \end{tabular}
\end{table}

\subsection{Laughlin states}
Firstly, we review the Laughlin states which are the most fundamental FQH states.\cite{laughlin_1983_Anomalous,haldane_1985_Periodic,jain_1989_Compositefermion,jain_1989_Incompressible,jain_1990_Theory,prange_1990_Quantum}
Among the FQH states induced by strong magnetic fields in two-dimensional electron systems, the Laughlin state is the simplest prototype, characterized by quasi-particle excitations obeying Abelian anyonic statistics.\cite{laughlin_1983_Anomalous,arovas_1984_Fractional,kivelson_1985_Consequences,wen_1995_Topological}
Serving as a critical reference point in both theoretical and experimental investigations of FQH systems, the Laughlin state warrants a systematic formulation of its bulk and edge theories.

\subsubsection{Bulk theory}
The filling factor becomes $ν=1/(2m+1)$ with integer $m$.
The Laughlin wave function $Ψ_\mathrm{Lau}$ is introduced as\cite{laughlin_1983_Anomalous,prange_1990_Quantum}
\begin{align}
  Ψ_\mathrm{Lau} = ∏_{i<j}(z_i - z_j)^{2m+1} \exp\qty[-\frac{1}{4l_B^2}\sum_{i}|z_i|^2] ,
\end{align}
where $z_{j}=x_{j}+iy_{j}$ is the complex coordinate of the $j$-th electron and $l_B$ is the magnetic length. 
This wave function yields an essentially lowest-energy state, reflecting strong interactions among electrons.
The quasi-holes obey anyonic statistics and carry fractionally quantized charge $e^{\ast} = e/(2m+1)$.\cite{laughlin_1983_Anomalous,wen_1995_Topological}

\subsubsection{Edge theory}
\label{sssec:Laughlin_edge_theory}
In FQH systems, incompressible fluids arise, suggesting that edge flows exist. 
This was demonstrated by Wen in the form of coupling with gauge fields.\cite{stone_1991_Vertex,wen_1991_EDGE,wen_1995_Topological,fradkin_2013_Field}
The action for the edge state of Laughlin states $S_\mathrm{Lau}$ is given by
\begin{align}
  S_{\mathrm{Lau}} = S_{\mathrm{boson}}
  = \int d\tau dx \left\{ \frac{1}{4\pi\nu} \partial_x\phi (-i\partial_\tau + v\partial_x)\phi \right\},
  \label{eq:Laughlin_edge_action}
\end{align}
where filling factor $ν = 1/(2m+1)$, $ϕ$ is the bosonic field which describes the edge excitations, and $v$ is the velocity of the bosonic field.
The commutation relation of the bosonic field is given by\cite{fradkin_2013_Field}
\begin{align}
  \comm{ϕ(x)}{ϕ(x')} = iπν\sgn(x-x'),
  \label{eq:commutation_relation}
\end{align}
and the fundamental quasi-hole and electron operators, summarized in \cref{tab:operators}, are given by
\begin{align}
  Ψ_\mathrm{QH}^\mathrm{Lau} = e^{iϕ}
  \qc{}
  Ψ_\mathrm{el}^\mathrm{Lau} = e^{iϕ/ν}.
\end{align}
Other quasi-particle operators carrying different fractional charges are summarized in Tab.~\ref{tab:quasi-particle_operators}.

\subsection{Moore--Read states}
The Moore--Read (MR) state was the first theoretical model to exhibit non-Abelian anyonic statistics within FQH effect context.\cite{moore_1991_Nonabelions}
Experimental signatures consistent with the Moore--Read phase have been reported at the $ν=5/2$ FQH plateau.\cite{willett_1987_Observation,greiter_1991_Paired}
This filling factor corresponds to two fully filled Landau levels ($ν=2$) plus a half-filled second Landau level ($ν=1/2$), which is widely believed to host the paired, non-Abelian Moore--Read state. 
However, the Moore--Read state faces competition from other candidate phases at $ν=5/2$, including the anti-Pfaffian and various Abelian states, and a conclusive identification remains an open question.\cite{levin_2007_ParticleHole,lee_2007_ParticleHole}
Distinct from conventional FQH states, the Moore--Read state is characterized by a richer structure of quasiparticle excitations and associated edge theories, resulting from its paired-state wave function.\cite{read_2000_Paired}
In the following, we review the bulk and edge theories of the Moore--Read state.

\subsubsection{Bulk theory}
The bulk Moore--Read wave function incorporates a pairing structure among electrons and is given by\cite{moore_1991_Nonabelions,read_2000_Paired,morf_1998_Transition}
\begin{align}
  Ψ_\mathrm{MR} = \Pf\biggl(\frac{1}{z_i - z_j}\biggr)∏_{i<j}(z_i - z_j)^{m} \exp\biggl[-\frac{1}{4l_B^2}\sum_{i}|z_i|^2\biggr],
\end{align}
where $\Pf(\hat{M})$ denotes the Pfaffian of an anti-symmetric matrix $\hat{M}$, and $m=2$ yields the non-Abelian state at filling fraction $ν=1/2$.
This wave function describes a lowest-energy state with strong electron correlations, and the pairing encoded by the Pfaffian leads to central charge $c=1/2$ conformal field theory (CFT) quasi-particle excitations.\cite{belavin_1984_Infinite,moore_1989_Classical,blok_1990_Effective,moore_1991_Nonabelions,francesco_2012_Conformal}
This is called Ising anyons and exhibit non-Abelian braiding statistics.
The quasi-holes also obey non-Abelian anyonic statistics and carry fractionally quantized charge $e^{\ast} = e/4$.\cite{nayak_1996_2nquasihole}

\subsubsection{Edge theory}
The edge theory of the Moore--Read state consists of a charged boson mode and a Majorana fermion mode.\cite{elitzur_1989_Remarks,milovanovic_1996_Edge}
The effective edge action $S_\mathrm{MR}$ reads
\begin{align}
  S_\mathrm{MR} 
  =&\; S_\mathrm{boson} + ∫ \dd{τ}\dd{x} \Biggl\{
    ½ψ\qty(\pdv{τ}+iv_{n}\pdv{x})ψ
  \Biggr\}\\
  =&\; S_\mathrm{boson} + S_{ψ} \notag,
  \label{eq:MR_edge_action}
\end{align}
where $S_\mathrm{boson}$ is the same form as \cref{eq:Laughlin_edge_action} includes the bosonic field $ϕ(x,t)$ describing the charge sector, $ψ(x,t)$ is the Majorana fermion field, and $v_{n}$ denotes the velocity of the Majorana fermion field.
The commutation relation of the bosonic field is the same as \cref{eq:commutation_relation}.
The fundamental quasi-hole and electron operators, summarized in Tab.~\ref{tab:operators}, are given by
\begin{align}
  Ψ_\mathrm{QH}^\mathrm{MR} = σ e^{iϕ/2}
  \qc{}
  Ψ_\mathrm{el}^\mathrm{MR} = ψ e^{iϕ/ν},
\end{align}
where $σ$ is the spin operator in the Ising CFT.
There is a bulk--edge correspondence\cite{witten_1989_Quantum,witten_1992_Holomorphic,wen_1990_Chiral,wen_1991_EDGE}: the structure and statistics of edge operators reflect the non-Abelian topological order and fusion rules of quasi-particles in the bulk Moore--Read state.
The other quasi-particle candidates, generated by these fusion rules and cataloged by their fractional charges, are summarized in Tab.~\ref{tab:quasi-particle_operators}.

\subsection{Read--Rezayi states}
The Read--Rezayi (RR) state generalizes the paired Moore--Read state to clustered states involving $k$-body correlations, and is one of the most prominent non-Abelian FQH phases.\cite{read_1999_Paired}
These states support quasi-particle excitations described by the $\mathbb{Z}_{k}$ parafermion CFT, featuring non-Abelian braiding statistics and richer topological structure.
The Read--Rezayi states, especially at $k=3$ corresponding to filling fraction $ν=3/5$, are of fundamental interest for both topological quantum computation and the theoretical understanding of clustered quantum Hall phases.
The Read--Rezayi $k=3$ state has been proposed as a candidate for the first excited Landau level at filling factor $\nu=13/5$.\cite{read_1999_Paired}
In the following, we formulate the bulk and edge theories of the Read--Rezayi state.

\subsubsection{Bulk theory}
The bulk wave function of the Read--Rezayi state is constructed via clustering of $k$ electrons and can be written as\cite{read_1999_Paired}
\begin{align}
  \Psi_\mathrm{RR} = &\mathcal{S} \left( \prod_{a=1}^k \prod_{i<j} (z_i^a - z_j^a)^2 \right) \notag\\
  &\times \prod_{i<j} (z_i - z_j) \exp \left[ -\frac{1}{4l_B^2} \sum_i |z_i|^2 \right],
\end{align}
where $\mathcal{S}$ is the symmetrization over all possible partitions of the electrons into clusters of $k$, and  the filling fraction $ν= k/(k+2)$ (for $k=3$, the filling is $ν=3/5$). 
This wave function encodes strong, multi-electron correlations and supports non-Abelian quasi-particles obeying $\mathbb{Z}_{k}$ parafermionic fusion rules.
The quasi-hole also obey non-Abelian anyonic statistics and carry fractionally quantized charge ($e^{\ast} = e/5$ for $k=3$).

\subsubsection{Edge theory}
The edge theory of the Read--Rezayi $k=3$ state consists of a chiral charge boson mode with $c=1$ and a parafermion sector described by the $\mathbb{Z}_{3}$ CFT with $c=4/5$. 
The effective action is\cite{law_2008_Probing,bishara_2008_Quantum,braggio_2012_Quasiparticle}
\begin{align}
  S_\mathrm{RR} 
  =& S_\mathrm{boson} + ∫ \dd{τ}\dd{x}\mathcal{L}_{\mathbb{Z}_3}\\
  =& S_\mathrm{boson} + S_{\mathbb{Z}_3} \notag,
  \label{eq:RR_edge_action}
\end{align}
where $\mathcal{L}_{\mathbb{Z}_3}$ is the action for the $\mathbb{Z}_3$ parafermion CFT.
The commutation relation of the bosonic field is the same as \cref{eq:commutation_relation}.
The fundamental quasi-hole and electron operators, summarized in Tab.~\ref{tab:operators}, are given by
\begin{align}
  Ψ_\mathrm{QH}^\mathrm{RR} = σ₁ e^{iϕ/3}
  \qc{}
  Ψ_\mathrm{el}^\mathrm{RR} = ψ₁ e^{iϕ/ν},
\end{align}
where $σ_1$ and $ψ_1$ are the primary fields in the $\mathbb{Z}_3$ parafermion CFT; here $σ_1$ denotes the spin and $ψ_1$ denotes the $\mathbb{Z}_3$ parafermion field.
The parafermion CFT sector gives the edge excitations their non-Abelian braiding and clustering properties, directly reflecting the bulk topological order and quasi-particle fusion rules.
A complete list of the quasi-particle candidates, cataloged by their fractional charges according to these fusion rules, is summarized in Tab.~\ref{tab:quasi-particle_operators}.

\subsection{Tunneling Hamiltonian of point contact geometry}
We consider the tunneling phenomena in various FQH states.\cite{kane_1992_Transmission,kane_1992_Transport,fendley_1995_Exact,dec.chamon_1997_Distinct,imura_1997_Quantum,furusaki_1993_Resonant,nomura_2001_Strong,fendley_2006_Dynamical,fendley_2007_Edge,fendley_2009_Boundary,ito_2012_QuasiParticle,ahmed_2026_Universal,fradkin_2013_Field}
The point-contact connects two counterpropagating chiral edges, which are described by independent holomorphic (right-moving) and antiholomorphic (left-moving) sectors as shown in \cref{fig: MR_Geometry}(a) and (b).
Globally the edge theory is therefore a full CFT obtained by combining these two chiral halves.
Local tunneling operators at the contact are constructed as products of right- and left-moving quasi-particle operators evaluated at $x=0$:
\begin{align}
  H_\alpha
  = \frac{\Gamma_{\alpha}}{2}\qty[\qty(\Psi^\mathrm{L}_\alpha)^\dagger(x=0) \Psi^\mathrm{R}_\alpha(x=0) + \qhc],
  \label{eq:tunneling_Hamiltonian}
\end{align}
where $\Gamma_{\alpha}$ is the amplitude of the tunneling process, and $\alpha$ is quasi-particle (QPT) or electron (ET).
The quasi-hole and electron operators for various FQH states are summarized in \cref{tab:operators}.
Following \cref{eq:tunneling_Hamiltonian}, and expressing the right- and left-moving chiral bosonic fields as $\phi^\mathrm{R}$ and $\phi^\mathrm{L}$, the electron and quasi-particle tunneling Hamiltonians are strictly given by the following forms, respectively:
\begin{align}
  H_\mathrm{ET\phantom{T}} 
    =&\; \frac{\Gamma_\mathrm{ET\phantom{T}}}{2} \biggl[ \eta(z)\bar{\eta}^\dagger(\bar{z}) \exp(i\frac{\phi^\mathrm{R}(x=0)-\phi^\mathrm{L}(x=0)}{\nu})\notag\\ &\hspace{20mm}+ \qhc \biggr],
    \label{eq:ET_Hamiltonian_full}\\
  H_\QPT 
    =&\; \frac{\Gamma_\QPT}{2} \biggl[ \zeta(z)\bar{\zeta}^\dagger(\bar{z}) \exp(i\frac{\phi^\mathrm{R}(x=0)-\phi^\mathrm{L}(x=0)}{n})\notag\\ &\hspace{20mm}+ \qhc \biggr],
    \label{eq:QPT_Hamiltonian_full}
\end{align}
where $\eta(z)$ and $\zeta(z)$ are the primary operators
\begin{align}
  \eta(z) = 
  \begin{cases}
    I \\
    \psi(z)\\
    \psi_1(z) 
  \end{cases}\hspace{-3mm},\;
  \zeta(z) = 
  \begin{cases}
    I\\
    \sigma(z)\\
    \sigma_1(z) 
  \end{cases}\hspace{-3mm}
  \begin{aligned}
    &\text{(Laughlin)}\\
    &\text{(Moore--Read)}\\
    &\text{(Read--Rezayi)}
  \end{aligned},
  \label{eq:zeta_eta_operator}
\end{align}
and $n=1,2,3$ for Laughlin, Moore--Read, and Read--Rezayi states, respectively.
For compactness, we denote these fields as $\eta(z)$ and $\bar{\eta}(\bar z)$ (and similarly for $\zeta$), where the presence or absence of a bar distinguishes the holomorphic (right-moving) and antiholomorphic (left-moving) edge sectors.

Here, we also discuss the full CFT structure of boson fields.\cite{fradkin_2013_Field}
The full bosonic action is $S_\mathrm{boson}^\mathrm{full} = S_\mathrm{boson}(\phi^\mathrm{R}) + S_\mathrm{boson}(\phi^\mathrm{L})$ which is defined in \cref{eq:Laughlin_edge_action}.
We introduce even and odd combinations of the bosonic fields as
\begin{align}
  \begin{split}
    \phi_\mathrm{o} =& \phi^\mathrm{R} - \phi^{\mathrm{L}},\\
    \phi_\mathrm{e} =& \qty(\phi^\mathrm{R} + \phi^{\mathrm{L}})/2,
  \end{split}
\end{align}
the action is decoupled into two sectors as
\begin{align}
  S_\mathrm{boson}^{\mathrm{full}}
  =& \int \dd{\tau}\dd{x} \frac{1}{8\pi\nu}\Biggl\{
    \partial_x\phi_\mathrm{o}\qty(
    -i\partial_\tau +v\partial_x
    )\phi_\mathrm{o}
  \Biggr\} \notag\\
  +& \int \dd{\tau}\dd{x} \frac{1}{2\pi\nu}\Biggl\{
    \partial_x\phi_\mathrm{e}\qty(
    -i\partial_\tau +v\partial_x
    )\phi_\mathrm{e}
    \Biggr\} \notag\\
  &= S_\mathrm{U(1),o} + S_\mathrm{U(1),e}.
  \label{eq:boson_action_even_odd}
\end{align}
While the even sector $\phi_\mathrm{e}$ is decoupled from the tunneling process, the odd sector $\phi_\mathrm{o}$ couples to the tunneling operators in \cref{eq:ET_Hamiltonian_full,eq:QPT_Hamiltonian_full}.
Therefore, we only consider the odd sector in the following discussion, and adopt the following notations for simplicity:
\begin{align}
  S_\mathrm{U(1)} \equiv S_\mathrm{U(1),o}\qc{} \phi \equiv \phi_\mathrm{o}.
  \label{eq:S_U1}
\end{align}
Consequently, since the primary fields are Hermitian in the Laughlin and Moore--Read states (i.e., $\eta^\dagger = \eta$ and $\zeta^\dagger = \zeta$), the tunneling Hamiltonians simply reduce to the following compact cosine forms:
\begin{align}
  H_\ET 
  =&\; \Gamma_\mathrm{ET\phantom{T}}\eta(z)\bar{\eta}(\bar{z}) \cos\qty(\frac{\phi}{ν}),
    \label{eq:ET_Hamiltonian}\\
  H_\QPT 
    =&\; \Gamma_\QPT \zeta(z)\bar{\zeta}(\bar{z}) \cos\qty(\frac{\phi}{n}).
    \label{eq:QPT_Hamiltonian}
\end{align}
For the Read--Rezayi state, however, the primary fields are not Hermitian ($\sigma_1^\dagger = \sigma_2 \neq \sigma_1$), and the Hamiltonians cannot be factorized into a single cosine term. The exact multi-branch potential structure for this state will be explicitly treated using matrix representations in \cref{sec:duality_RR}.

\section{Duality for Laughlin states} 
\label{sec:duality_Laughlin}
 We would like to introduce the framework of duality.\cite{schmid_1983_Diffusion,fisher_1985_Quantum,callan_1992_Phase,kane_1992_Transmission,imura_1997_Quantum,furusaki_1993_Resonant} 
First of all, we review the duality for Laughlin states with techniques by Caldeira-Leggett model and instanton method.

\subsection{Partition function with tunneling}
We consider the quasi-particle tunneling process at the point contact.
The partition function is given by
\begin{align}
  Z 
  =& ∫ \DD{ϕ} \exp\biggl[
    - S_\mathrm{U(1)}
     + ∫\dd{τ}Γ_\QPT\cos(ϕ(x=0)) 
  \biggr]\\
  =&∫\DD{ϕ} \exp\left[-S_\mathrm{U(1)+QPT}\right].
  \label{eq:Laughlin_Z_U1+QPT}
\end{align}
We map the partition function to the Caldeira-Leggett model\cite{caldeira_1981_Influence} as shown in \cref{appeq:Laughlin_eff_action} in Appendix \ref{app:CL_mapping}.
The point contact action $S_\mathrm{U(1)+QPT}$ is rewritten as the effective action for the phase variable $θ(τ)$ as
\begin{align}
  S_\mathrm{U(1)+QPT}\qty[\theta]
  = \sum_{ω}\frac{|ω|}{4πν}θ(-ω)θ(ω)
    -∫ \dd{τ}Γ_\QPT\cos(θ(τ)).
    \label{eq:Laughlin_eff_action}
\end{align}
In the view point of the dynamics for $θ$, the first term is the friction term in the Caldeira-Leggett theory, and the second term is regarded as the periodic potential with large amplitude $Γ_\QPT$ in the strong coupling regime.

\subsection{Instanton-gas approximation}
\label{ssec:Laughlin_instanton}
\begin{figure}[htbp]
    \centering
    \begin{minipage}[t]{\linewidth}
      \begin{tikzpicture}[scale=1.1,samples=300]
        \draw[->,>=stealth,semithick] (-3,0)--(3,0) node[right]{$θ$}; 
        \draw[->,>=stealth,semithick] (0,-1)--(0,1) node[left]{$ $}; 
        \draw (-3*pi/4,0) -- (-3*pi/4,-0.07);
        \draw (-2*pi/4,0) -- (-2*pi/4,-0.07);
        \draw (-pi/4,0) -- (-pi/4,-0.07);
        \node [anchor=north east,font=\scriptsize] at (0.05,0.03) {$0$};
        \draw (pi/4,0) -- (pi/4,-0.07);
        \node [anchor=north,font=\scriptsize] at (pi/4,0.03) {$π$};
        \draw (2*pi/4,0) -- (2*pi/4,-0.07);
        \node [anchor=north,font=\scriptsize] at (2*pi/4,0.03) {$2π$};
        \draw (3*pi/4,0) -- (3*pi/4,-0.07);
        \draw[thick,domain=-3:3] plot(\x,{-0.5*cos(4*\x r)});

        \draw[dashed] (0,0.5) -- (pi/4,0.5);
        \node [anchor=south west,font=\scriptsize] at (0,0.5) {$Γ_\QPT$};

        \draw[thick,red,dotted] (-pi/2,-0.5) circle[radius=0.1];
        \fill[red] (0,-0.5) circle[radius=0.1];
        \draw[thick,red] (+pi/2,-0.5) circle[radius=0.1];

        \draw[red,->] (0,-0.6) to[in=-120,out=-60] (pi/2,-0.6);
        \node [red,anchor=north,font=\scriptsize] at (pi/4,-1) {instanton};
        \node [red,anchor=south,font=\scriptsize] at (pi/4,-1.0) {$e_i=+1$};
        \draw[red,->,dotted] (0,-0.6) to[in=-60,out=-120] (-pi/2,-0.6);
        \node [red,anchor=south,font=\scriptsize] at (-pi/4,-1.0) {$e_i=-1$};
        \node [red,anchor=north,font=\scriptsize] at (-pi/4-0.1,-1) {anti-instanton};
      \end{tikzpicture}
    \end{minipage}
    \caption{
      Schematic picture of the instanton and anti-instanton solutions in the periodic potential of \cref{eq:Laughlin_eff_action}.
      Solid (dotted) arrow represents the (anti-)instanton solution. 
    }
    \label{fig: instanton}
  \end{figure}
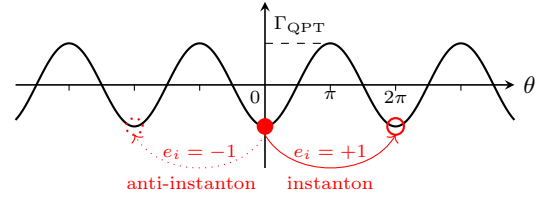

  The excitations for the strong coupling limit are described by the instanton solutions in the same way as the problems of single impurity in non-chiral Luttinger liquid.
  Using the instanton method we construct an effective action for the strong coupling regime. 
  We introduce a function defined as
  \begin{align}
    \dv{θ_\mathrm{ins}}{τ} ≡ h(τ),
    \label{eq:Laughlin_dtheta_ins}
  \end{align}
  where $θ_\mathrm{ins}$ is the solution for single instanton and $h(\tau)$ is a sudden phase jump.
  So the differential of the $n$ instanton solution and its Fourier coefficient are written as
  \begin{align}
    \dv{θ_n}{\tau} =& \sum_{i=1}^{n}e_i h(\tau-\tau_i)\\
    -iωθ_n(\omega) =& \sum_{i=1}^{n} e_i\tilde{h}(\omega)e^{i\omega\tau_i},
  \end{align}
  where $eᵢ$ is the charge of the instantons, and $\tilde{h}(ω)$ is the Fourier coefficient of $h(τ)$.
  Especially, a difference of $θ$ in the instanton becomes $2π$ as shown in \cref{fig: instanton}. 
  Thus the value is as follows:
  \begin{align}
    \tilde{h}(ω=0) = \frac{1}{\sqrt{β}}∫\dd{τ}h(τ)=\frac{2π}{\sqrt{β}}.
  \end{align} 

   In the strong-coupling limit (large $\Gamma_\mathrm{QPT}$), the field $\theta$ is strongly pinned to one of the minima of the periodic potential in \cref{eq:Laughlin_eff_action}.
   The low-energy dynamics are then dominated by instanton events, which are classical solutions transitioning between these adjacent minima.
   For an $n$-instanton configuration $\theta_n(\tau)$, the potential term is approximately constant outside the instanton cores, and its contribution during the phase slip is absorbed into the fugacity of a single instanton.
   Thus, the effective action for the $n$-instanton configuration, $S_\mathrm{ins}^{(n)}$, is predominantly determined by the friction term.
   Substituting the $n$-instanton solution $\theta_n$ into the friction term, and noting that the duration of each instanton event is much shorter than the mean interval between them, we can safely evaluate the interaction in the low-frequency limit ($\omega \to 0$).
   Under this approximation, we can neglect the $\omega$-dependence of $\tilde{h}(\omega)$ and evaluate the action as
  \begin{align}
    S_\mathrm{ins}^{(n)}[\theta_n]
    \equiv&\; \sum_{\omega}\frac{|\omega|}{4\pi\nu}\theta_n(-\omega)\theta_n(\omega)\notag\\
    =&\; \sum_{\omega}\frac{|\omega|}{4\pi\nu}\qty(\frac{1}{i\omega}\sum_{i=1}^n e_i\tilde{h}(\omega)e^{i\omega\tau_i})\notag\\
    &\;\;\;\times\qty(\frac{1}{-i\omega}\sum_{j=1}^n e_j\tilde{h}(-\omega)e^{-i\omega\tau_j})\notag\\
    =&\; \sum_{i,j}^n\sum_{\omega}\frac{1}{4\pi\nu|\omega|}e_i e_j \tilde{h}(-\omega)\tilde{h}(\omega)e^{i\omega(\tau_i-\tau_j)}\notag\\
    \simeq&\; \sum_{i,j}^n\sum_{\omega}\frac{1}{4\pi\nu|\omega|}\qty(\tilde{h}(\omega=0)\tilde{h}(\omega=0))e^{i\omega(\tau_i-\tau_j)}e_i e_j\notag\\
    =&\; \sum_{i,j}^n\qty( \frac{\pi}{\nu}\frac{1}{\beta}\sum_{\omega}\frac{1}{|\omega|}e^{i\omega(\tau_i-\tau_j)} )e_i e_j.
    \label{eq:Laughlin_S_ins}
  \end{align}
  By summing over all possible instanton configurations (the number of instantons $n$, their positions $\tau_i$, and their charges $e_i$), the grand-canonical partition function is given as
  \begin{align}
    Z_\mathrm{ins}
    =& \sum_{n=0}^{\infty}\sum_{\{e_i\}}\frac{1}{n!}\int_0^\beta\dd{\tau_n}\int_0^\beta\dd{\tau_{n-1}}\cdots\int_0^\beta\dd{\tau_1} \prod_{\ell=0}^n z_0(\tau_\ell) \notag\\
     & \times \exp\qty[ -\sum_{i,j}^n e_i e_j \frac{\pi}{\nu}\frac{1}{\beta}\sum_{\omega}\frac{e^{-i\omega(\tau_i-\tau_j)}}{|\omega|} ],
     \label{eq:instanton_gas_partition_function}
  \end{align}
  where $z_0 \propto e^{-S_\mathrm{single}}$ is the instanton fugacity, with $S_\mathrm{single}$ being the classical action of a single instanton core corresponding to the solution in \cref{eq:Laughlin_dtheta_ins}.
  We here consider the grand canonical ensemble, so that the fugacity appears.
  We can stress that its choice has some arbitrariness but it is important to establish the duality.
  The factor $z_0$ is much smaller than the value of $\Gamma_\mathrm{QPT}$ because the instanton probability is suppressed by the large potential amplitude $\Gamma_\mathrm{QPT}$.

  Finally, we introduce Stratonovich--Hubbard field $Θ$ to rewrite the dual action as
  \begin{align}
    S_\mathrm{dual}[Θ]
    =& \sum_{ω}\frac{|ω|}{4πν}Θ(-ω)Θ(ω)
      -∫ \dd{τ}2z₀\cos(\frac{Θ(τ)}{ν})
  \label{eq:Laughlin_eff_action_ET}
  \end{align}
  which is derived as introduced in \cref{appeq:Laughlin_eff_action_Theta} in Appendix \ref{app:SH_transformation}.
  This is the partition function for the electron tunneling process in the weak coupling regime.
  Thus, we find the duality between the quasi-particle tunneling in the weak coupling regime and the electron tunneling in the strong coupling regime.
  As illustrated in \cref{fig: MR_Geometry}(c), when quasi-particle tunneling becomes strong the chiral current on the upper edge is redirected to the lower edge, effectively reconnecting the edges.
  This produces the configuration described by weak electron tunneling, so that the strong-coupling quasi-particle tunneling problem is mapped onto the weak-coupling electron tunneling problem.

\section{Duality for Moore--Read states with phase-shift} 
\label{sec:duality_MR}
We consider the strong coupling quasi-particle tunneling process at the point contact in Moore--Read states by introducing a phase-shift instanton.
\cite{nomura_2001_Strong,ito_2012_QuasiParticle,ahmed_2026_Universal}

\subsection{Partition function with tunneling}
  The partition function for Moore--Read state case is given by
  \begin{align}
    Z 
    =& Z_\mathrm{Ising}\Biggl<∫ \DD{ϕ} \exp\biggl[-S_\mathrm{U(1)} \notag\\
      &+ ∫\dd{τ}Γ_\QPT σ\bar{σ}\cos(\frac{ϕ(x=0)}{2}) 
    \biggr]\Biggr>_\mathrm{Ising-CFT}\notag\\
    =& Z_\mathrm{Ising}\Biggl<∫\DD{ϕ} \exp\Biggl[-S_\mathrm{U(1) +QPT} \Biggr]\Biggr>_\mathrm{Ising-CFT},
    \label{eq:MR_Z_U1+Ising+QPT}
  \end{align}
  where the quasi-particle tunneling term is given by \cref{eq:QPT_Hamiltonian} with \cref{eq:zeta_eta_operator} for Moore--Read states.
  Here, $Z_\mathrm{Ising}$ is the partition function of the Ising CFT, and $\expval{\cdots}_\mathrm{Ising-CFT}$ denotes the expectation value evaluated within this CFT framework.
  We map the partition function to the Caldeira-Leggett model as shown in \cref{appeq:Laughlin_eff_action} in Appendix \ref{app:CL_mapping}.
  The action for the phase variable $θ(τ)$ is given by
  \begin{multline}
    S_\mathrm{U(1)+QPT}[\theta,\sigma,\bar{\sigma}]
    = \sum_{ω}\frac{|ω|}{4πν}θ(-ω)θ(ω)\\
      -∫ \dd{τ}Γ_\QPT σ\bar{σ}\cos(\frac{θ(τ)}{2}).
    \label{eq:MR_eff_action}
  \end{multline}
  This action is almost the same as \cref{eq:Laughlin_eff_action} except for the factor $σ\bar{σ}$ and the coefficient of $θ$ in the cosine term.

  \subsection{Phase-shift instanton-gas approximation}
\begin{figure}[htb]
  \centering
    \begin{tikzpicture}[scale=0.8,samples=300]
    \begin{scope}[shift={(0,0)}]
      \begin{scope}[shift={(0,4)}]
        \node [anchor=west] at (-2.2,1.5) {(a) Moore--Read state case};

        \draw[->,>=stealth,semithick] (-2,0)--(2,0) node[right]{$θ/2$}; 
        \draw[->,>=stealth,semithick] (0,-1)--(0,1) node[left]{$ $}; 
        \draw[thick,domain=-1.8:1.8] plot(\x,{-0.5*cos(4*\x r)});
        \draw (-2*pi/4,0) -- (-2*pi/4,-0.07);
        \draw (-pi/4,0) -- (-pi/4,-0.07);
        \node [anchor=north east,font=\scriptsize] at (0.1,0.06) {$0$};
        \node [anchor=north,font=\scriptsize] at (pi/4,0.06) {$π$};
        \draw (pi/4,0) -- (pi/4,-0.07);
        \node [anchor=north,font=\scriptsize] at (2*pi/4,0.06) {$2π$};
        \draw (2*pi/4,0) -- (2*pi/4,-0.07);

        \fill[red] (0,-0.5) circle[radius=0.1];
      \end{scope}

      \begin{scope}[shift={(0,2)}]
        \draw[teal,->,>=stealth] (0.0,0.5) to[in=60,out=-60] (0.0,-0.5); 
        \node [teal,anchor=west,font=\Large] at (0.2,0.0) {$π$ shift};
      \end{scope}
      
      \begin{scope}[shift={(0,0)}]
        \draw[->,>=stealth,semithick] (-2,0)--(2,0) node[right]{$θ/2$}; 
        \draw[->,>=stealth,semithick] (0,-1)--(0,1) node[left]{$ $}; 
        \draw[thick,domain=-1.8:1.8] plot(\x,{-0.5*cos(4*(\x-pi/4) r)});
        \draw[thin,domain=-1.8:1.8,dotted] plot(\x,{-0.5*cos(4*\x r)});
        \draw (-2*pi/4,0) -- (-2*pi/4,-0.07);
        \draw (-pi/4,0) -- (-pi/4,-0.07);
        \node [anchor=north east,font=\scriptsize] at (0.1,0.06) {$0$};
        \node [anchor=north,font=\scriptsize] at (pi/4,0.06) {$π$};
        \draw (pi/4,0) -- (pi/4,-0.07);
        \node [anchor=north,font=\scriptsize] at (2*pi/4,0.06) {$2π$};
        \draw (2*pi/4,0) -- (2*pi/4,-0.07);

        \draw[red,thick,dotted] (0,-0.5) circle[radius=0.1];
        \fill[red] (pi/4,-0.5) circle[radius=0.1];

        \draw[red,->] (0,-0.6) to[in=-120,out=-60] (pi/4,-0.6);
        \node [red,anchor=north,font=\normalfont] at (pi/8,-1) {$π$-shift instanton};
      \end{scope}
    \end{scope}

    \begin{scope}[shift={(3.1,0)}]
      \draw[thick,dotted] (0.0,-2.0) -- (0.0,6.0);
    \end{scope}

    \begin{scope}[shift={(5.5,0)}]
      \begin{scope}[shift={(0,4)}]
        \node [anchor=west] at (-2.2,1.5) {(b) Read--Rezayi state case};

        \draw[->,>=stealth,semithick] (-2,0)--(2,0) node[right]{$θ/3$}; 
        \draw[->,>=stealth,semithick] (0,-1)--(0,1) node[left]{$ $}; 
        \draw[thick,domain=-1.8:1.8] plot(\x,{-0.5*cos(4*\x r)});
        \node [anchor=north east,font=\scriptsize] at (0.1,0.06) {$0$};
        \draw (-2*pi/4,0) -- (-2*pi/4,-0.07);
        \draw (-pi/4,0) -- (-pi/4,-0.07);
        \node [anchor=north,font=\scriptsize] at (pi/4,0.06) {$π$};
        \draw (pi/4,0) -- (pi/4,-0.07);
        \node [anchor=north,font=\scriptsize] at (2*pi/4,0.06) {$2π$};
        \draw (2*pi/4,0) -- (2*pi/4,-0.07);

        \fill[red] (0,-0.5) circle[radius=0.1];
      \end{scope}

      \begin{scope}[shift={(0,2)}]
        \draw[teal,->,>=stealth] (0,0.5) to[in=60,out=-60] (0.0,-0.5); 
        \node [teal,anchor=west,font=\large] at (0.2,0.0) {$\dfrac{2}{3}π$ shift};
      \end{scope}

      \begin{scope}[shift={(0,0)}]
        \draw[->,>=stealth,semithick] (-2,0)--(2,0) node[right]{$θ/3$}; 
        \draw[->,>=stealth,semithick] (0,-1)--(0,1) node[left]{$ $}; 
        \draw[thick,domain=-1.8:1.8] plot(\x,{-0.5*cos(4*(\x-2*pi/12) r)});
        \draw[thin,domain=-1.8:1.8,dotted] plot(\x,{-0.5*cos(4*(\x) r)});
        \draw (-2*pi/4,0) -- (-2*pi/4,-0.07);
        \draw (-pi/4,0) -- (-pi/4,-0.07);
        \node [anchor=north east,font=\scriptsize] at (0.1,0.06) {$0$};
        \node [anchor=south,font=\scriptsize] at (pi/4,-0.06) {$π$};
        \draw (pi/4,0) -- (pi/4,-0.07);
        \node [anchor=north,font=\scriptsize] at (2*pi/4,0.06) {$2π$};
        \draw (2*pi/4,0) -- (2*pi/4,-0.07);

        \draw[red,thick,dotted] (0,-0.5) circle[radius=0.1];
        \fill[red] (2*pi/12,-0.5) circle[radius=0.1];

        \draw[red,->] (0,-0.6) to[in=-120,out=-60] (2*pi/12,-0.6);
        \node [red,anchor=north,font=\normalfont] at (pi/8,-1) {$\frac{2}{3}π$-shift instanton};
      \end{scope}


    \end{scope}
    \begin{scope}[shift={(3,-2.6)}]
      \node [anchor=center,font=\large] at (0.0,0.0) {\textit{Phase-shift instanton}};
        \draw[,->] (-2.5,0.0) to[in=-90,out=180] (-3.5,1.0);
        \draw[,->] (+2.5,0.0) to[in=-90,out=  0] (+3.5,1.0);
    \end{scope}

    \end{tikzpicture}
  \caption{
    Schematic picture of the phase-shift instanton process.
    We define ''\textit{phase-shift instanton}'' as the instanton process that occurs with a phase-shift of the potential.
    Upper figure shows the potential of \cref{eq:MR_eff_action,eq:RR_eff_action}.
    Lower figure represents the details of the phase-shift instanton process in each case.
    Solid (dotted) line represents the potential after (before) the instanton process.
    (a) shows the Moore--Read state case with $π$-shift instanton, and (b) shows the Read--Rezayi state case with $2π/3$-shift instanton.
  }
  \label{fig:phase-shift_instanton}
\end{figure}
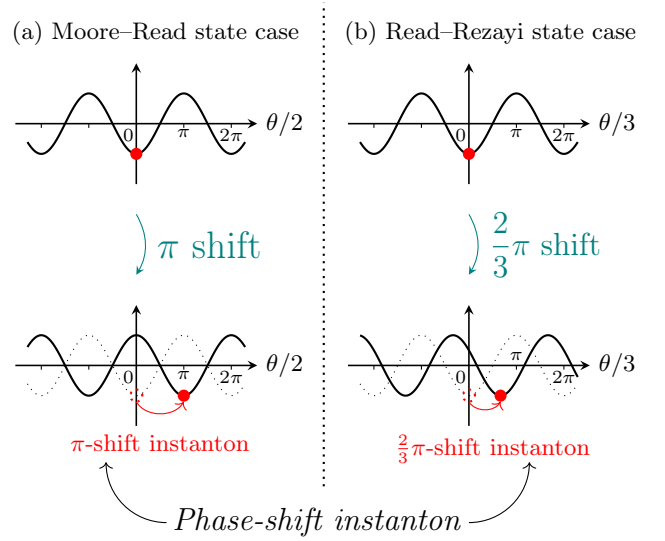

  We consider the instanton method to derive the effective action for the strong coupling regime.
  We introduce the instanton solution the same as the Laughlin state case as shown in \cref{fig: instanton}.
  However, in this case, the shift of the field $θ$ becomes $4π$ because of a factor of $1/2$ in the argument of cosine term in \cref{eq:MR_eff_action}.
  Therefore, we cannot derive the dual action the same as the Laughlin state case directly.
  To resolve this issue, we focus on the factor $\sigma\bar{\sigma}$ in the tunneling term.
  Following the mapping to the boundary impurity problem,\cite{fendley_2006_Dynamical,fendley_2007_Edge} the expectation value over the topological CFT sector can be effectively evaluated by replacing the local product of the primary fields with a finite-dimensional impurity spin matrix.
  Specifically, the primary field product $\sigma\bar{\sigma}$ corresponds to the $1/2$-spin operator $\hat{\sigma}$ in the Kondo problem\cite{fendley_2006_Dynamical, fendley_2007_Edge}:
  \begin{align}
    \hat{σ} = \mqty(
      1 & 0\\
      0 & -1
    )
     = \mqty(
      1 & 0\\
      0 & e^{iπ}
    ).
  \end{align}
  Using this mapping, we can describe the effect of $σ\bar{σ}$ as the $π$ phase-shift of the potential for the instanton process as shown in \cref{fig:phase-shift_instanton}(a).
  We will refer to the instanton that experiences a phase-shift due to the influence of spin operators as ''\textit{phase-shift instanton}.''
  Therefore, the phase-shift instanton process gets the shift of the field $θ$ by $2π$
  \begin{align}
    \tilde{h}(ω=0) = \frac{1}{\sqrt{β}}∫\dd{τ}h(τ)=\frac{2π}{\sqrt{β}}.
  \end{align} 
  To consider the grand-canonical ensemble of the phase-shift instantons, we can derive the effective action for the strong coupling regime as follow:
  \begin{align}
    Z_\mathrm{ins} 
    =& \Biggl<\sum_{n=0}^{∞}\sum_{\qty{eᵢ}}
    \frac{1}{n!}
    ∫_{0}^{β}\dd{τₙ}∫_{0}^{β}\dd{τ_{n-1}} ⋯ \notag\\
    \;& ∫_{0}^{β}\dd{τ₁}
    \prod_{ℓ=0}^{n} z₀\qty(τ_{\ell},\qty{\mathcal{O}},\qty{\bar{\mathcal{O}}})\notag\\
    ×
    &\exp\qty[
      -\,\sum_{ij}^{n}
      e_i e_j
      \frac{π}{ν}\frac{1}{β}∑_{ω}\frac{e^{-iω(τ_i-τ_j)}}{|ω|}
    ]\Biggr>_\mathrm{Ising-CFT},
    \label{eq:instanton_gas_partition_function_MR}
  \end{align}
  where $\mathcal{O}$ is an arbitrary primary field in the Ising CFT, and $\{\mathcal{O}\}$ denotes the corresponding conformal family generated by $\mathcal{O}$.
  Note that, since the specific tunneling primary field $\sigma\bar{\sigma}$ is replaced by the finite-dimensional matrix $\hat{\sigma}$, evaluating its expectation value reduces simply to taking a matrix trace.
  Since the fugacity $z_0$ physically represents the statistical weight of a single instanton event evaluated within the full CFT expectation value, it can generally incorporate internal degrees of freedom belonging to arbitrary conformal families.
  Consequently, the instanton fugacity incorporates these degrees of freedom, taking the form $z_0(\tau, \{O\}, \{\bar{O}\})$.
  While the specific primary field $\mathcal{O}$ is undetermined at this stage, the dual topological charge will restrict it to a specific primary field $\eta$, as we will show below.
  Finally, we introduce Stratonovich--Hubbard field $Θ$ to rewrite the dual action as
  \begin{align}
    &S_\mathrm{dual}\qty[Θ,\qty{\mathcal{O}},\qty{\bar{\mathcal{O}}}]
    = \sum_{ω}\frac{|ω|}{4πν}Θ(-ω)Θ(ω)\notag\\
     &\hspace{15mm} -∫ \dd{τ}2z_0\qty(\tau,\qty{\mathcal{O}},\qty{\bar{\mathcal{O}}})\cos(\frac{Θ(τ)}{ν})
    \label{eq:MR_eff_action_dual}
  \end{align}
    which is derived in Appendix \ref{app:SH_transformation}.
    Since the coefficient of $\Theta$ in the cosine term indicates tunneling of a particle with charge $e$, the conformal family $\qty{\mathcal{O}}$ in the fugacity is restricted to the candidates carrying charge $e$.
    There are two such candidates for the primary field, namely $\mathcal{O} = \psi$ and $\mathcal{O} = I$, which yield $z_0(\tau, \{\mathcal{O}\}, \{\bar{\mathcal{O}}\}) = w_0 \psi\bar{\psi}$ and $w_0 I\bar{I}$, respectively; see \cref{tab:quasi-particle_operators}.
    The coefficient $w_0$ represents the component independent of the conformal family.
    In particular, the dual action with $\mathcal{O} = I$ fails to reproduce quasi-hole tunneling under the inverse duality transformation. 
    Indeed, if we formally substitute $\nu = 1/2$ into \cref{eq:Laughlin_eff_action_ET} and apply the transformation, it would correspond to being mapped to $\cos(\theta)$, which does not correspond to the quasi-hole charge. 
    Although we have not provided a rigorous proof, this observation suggests that the candidate $\mathcal{O} = I$ can be ruled out. 
    Therefore, we can uniquely identify the true electron operator's primary field as $\eta = \psi$.
    We thus write $z_0(\tau, \{\mathcal{O}\}, \{\bar{\mathcal{O}}\}) = w_0 \eta\bar{\eta}$, with $w_0$ effectively serving as the bare electron tunneling amplitude $\Gamma_\mathrm{ET}$ in the dual regime.

  Therefore, in the dual picture, we conclude that
  \begin{align}
    S_\mathrm{dual}[Θ,\psi,\bar{\psi}]
    =& \sum_{ω}\frac{|ω|}{4πν}Θ(-ω)Θ(ω)\notag\\
     &\hspace{10mm} -∫ \dd{τ}2w₀ψ\bar{ψ}\cos(\frac{Θ(τ)}{ν}).
    \label{eq:MR_eff_action_ET}
  \end{align}
  This is the action for the electron tunneling process in the weak coupling regime.
  Thus, we find the duality between the quasi-particle tunneling in the strong coupling regime and the electron tunneling in the weak coupling regime even for Moore--Read states by considering the phase-shift instanton process as shown in \cref{fig:phase-shift_instanton}(a).

\section{Duality for Read--Rezayi states with phase-shift} 
\label{sec:duality_RR}
In this section, we consider the strong coupling quasi-particle tunneling process at the point contact in Read--Rezayi states.

\subsection{Partition function with tunneling}
  The partition function for Read--Rezayi $k=3$ state case is given by
  \begin{align}
    Z 
    =& Z_{\mathbb{Z}_3} 
    \Biggl<\int \DD{\phi} \exp\Biggl[
      -S_\mathrm{U(1)}\notag \\
      &\hspace{-2mm}+ \int\dd{\tau}\frac{\Gamma_\QPT}{2}\qty( \sigma_1\bar{\sigma}_1^\dagger e^{i\phi(x=0)/3} + \qhc ) 
    \Biggr]\Biggr>_{\mathbb{Z}_3\mathrm{-CFT}}\notag\\
    =& Z_{\mathbb{Z}_{3}}\Biggl<\int \DD{\phi} \exp\qty[-S_\mathrm{U(1) +QPT} ]\Biggr>_{\mathbb{Z}_3\mathrm{-CFT}},
    \label{eq:RR_Z_U1+Z3+QPT}
  \end{align}
  where the quasi-particle tunneling term is given by \cref{eq:QPT_Hamiltonian_full} with \cref{eq:zeta_eta_operator} for Read--Rezayi states.
  Here, $Z_{\mathbb{Z}_3}$ is the partition function of the $\mathbb{Z}_3$ parafermion CFT, and $\expval{\cdots}_{\mathbb{Z}_3\mathrm{-CFT}}$ denotes the expectation value evaluated within this CFT framework.
  We map the partition function to the Caldeira-Leggett model as shown in \cref{appeq:Laughlin_eff_action} in Appendix \ref{app:CL_mapping}.
  The action for the phase variable $θ(τ)$ is given by
  \begin{multline}
    S_\mathrm{U(1)+QPT}[\theta,\sigma_1,\bar{\sigma}_1]
    = \sum_{ω}\frac{|ω|}{4πν}θ(-ω)θ(ω)\\
      -∫ \dd{τ}\frac{Γ_\QPT}{2}\qty( σ₁\bar{σ}₁^{\dagger} e^{iθ(τ)/3}+\qty(σ₁\bar{σ}₁^{\dagger})^{\dagger} e^{-iθ(τ)/3}).
    \label{eq:RR_eff_action}
  \end{multline}
  This action is almost the same as \cref{eq:Laughlin_eff_action} except for the factor $σ₁\bar{σ}₁$ and a factor of $1/3$  in the argument of exponential term in \cref{eq:RR_eff_action}.

\subsection{Phase-shift instanton-gas approximation}
  We consider the instanton method to derive the effective action for the strong coupling regime.
  We also introduce the phase-shift instanton solution the same as the Moore--Read state case as shown in \cref{fig: instanton}.
  Here, the $\mathbb{Z}_3$ parafermion CFT is the critical theory of the three-state Potts model.\cite{dotsenko_1984_Critical}
  Following the mapping to the boundary impurity problem,\cite{fendley_2006_Dynamical,fendley_2007_Edge} the expectation value over the topological CFT sector can be effectively evaluated by replacing the local product of the primary fields with a finite-dimensional impurity spin matrix.
  Specifically, the non-Hermitian product $\sigma_1\bar{\sigma}_1^\dagger$ corresponds to the spin operator $\hat{\sigma}_1$ in the Potts model\cite{mong_2014_Parafermionic}:
  \begin{align}
    \hat{\sigma}_1 = \mqty(
      1 & 0 & 0 \\
      0 & e^{i2\pi/3} & 0 \\
      0 & 0 & e^{i4\pi/3}
    ).
  \end{align}
  By substituting this matrix representation into \cref{eq:RR_eff_action}, the tunneling term is explicitly rewritten as
  \begin{align}
    &\frac{1}{2}\qty( \sigma_1\bar{\sigma}_1^{\dagger} e^{i\theta/3} + \qty(\sigma_1\bar{\sigma}_1^{\dagger})^\dagger e^{-i\theta/3} ) \notag\\
    &\hspace{10mm}\to\frac{1}{2}\qty( \hat{\sigma}_1 e^{i\theta/3} + \hat{\sigma}_1^\dagger e^{-i\theta/3} ) \notag\\
    &\hspace{10mm}
    = \mqty(
      \cos\qty(\frac{\theta}{3}) & 0 & 0 \\
      0 & \cos\qty(\frac{\theta}{3}+\frac{2\pi}{3}) & 0 \\
      0 & 0 & \cos\qty(\frac{\theta}{3}+\frac{4\pi}{3})
    ).
  \end{align}
  This matrix explicitly shows that the internal state $\hat{\sigma}_1$ induces exactly a $2\pi/3$ phase-shift in the periodic potential.

  Therefore, the phase-shift instanton process gets the shift of the field $\theta$ by $2\pi$,
  \begin{align}
    \tilde{h}(ω=0) = \frac{1}{\sqrt{β}}∫\dd{τ}h(τ)=\frac{2π}{\sqrt{β}}.
  \end{align} 
  To consider the grand-canonical ensemble of the phase-shift instantons, we can derive the effective action for the strong coupling regime as follow:
  \begin{align}
    Z_\mathrm{ins} 
    =& \Biggl<\sum_{n=0}^{∞}\sum_{\qty{eᵢ}}
    \frac{1}{n!}
    ∫_{0}^{β}\dd{τₙ}∫_{0}^{β}\dd{τ_{n-1}} ⋯ \notag\\
    \;& ∫_{0}^{β}\dd{τ₁}
    \prod_{ℓ=0}^{n} z₀\qty(τ_{\ell},\qty{\mathcal{O}},\qty{\bar{\mathcal{O}}})\notag\\
    ×
    &\exp\qty[
      -\,\sum_{ij}^{n}
      e_i e_j
      \frac{π}{ν}\frac{1}{β}∑_{ω}\frac{e^{-iω(τ_i-τ_j)}}{|ω|}
    ]\Biggr>_{\mathbb{Z}_3\mathrm{-CFT}},
    \label{eq:instanton_gas_partition_function_RR}
  \end{align}
  where $\mathcal{O}$ is a primary field in the $\mathbb{Z}_3$ parafermion CFT.
  Note that, since the specific tunneling primary field $\sigma_1\bar{\sigma}_1^\dagger$ is replaced by the finite-dimensional matrix $\hat{\sigma}_1$, evaluating its expectation value reduces simply to taking a matrix trace. 
  Finally, we introduce the Stratonovich–Hubbard field $\Theta$ to rewrite the dual action as
  \begin{align}
    &S_\mathrm{dual}\qty[Θ, \qty{\mathcal{O}}, \qty{\bar{\mathcal{O}}^\dagger}]
    = \sum_{ω}\frac{|ω|}{4πν}Θ(-ω)Θ(ω)\notag\\
     &\hspace{10mm} -∫ \dd{τ} \qty[ z_0\qty(\tau, \qty{\mathcal{O}}, \qty{\bar{\mathcal{O}}^\dagger}) \exp(i\frac{Θ(τ)}{ν}) + \qhc ],
    \label{eq:RR_eff_action_ET_general}
  \end{align}
  which is derived in \cref{appeq:Laughlin_eff_action_Theta} in Appendix \ref{app:SH_transformation}.
  Since the coefficient of $\Theta$ in the exponential term indicates tunneling of a particle with charge $e$, the conformal family $\qty{\mathcal{O}}$ in the fugacity is restricted to the candidates carrying charge $e$.
  There are two such candidates for the primary field, namely $\mathcal{O} = \psi_1$ and $\mathcal{O} = \sigma_2$, which yield $z_0(\tau, \{\mathcal{O}\}, \{\bar{\mathcal{O}}^{\dagger}\}) = w_0 \psi_1\bar{\psi}_1^{\dagger}$ and $w_0 \sigma_2\bar{\sigma}_2^{\dagger}$, respectively; see \cref{tab:quasi-particle_operators}.\cite{law_2008_Probing,mong_2014_Universal}
  The phase-shift instanton gas method alone cannot distinguish between these two possibilities.
  However, in the strong coupling regime, the particle tunneling completely across the vacuum gap must physically be a true fermion.
  As we will verify in \cref{sec:conductance}, the $\psi_1$ channel precisely satisfies this fermionic requirement, whereas the $\sigma_2$ channel yields an anyonic dimension, which is physically prohibited from crossing the vacuum.
  Thus, physical principles uniquely select the true electron operator's primary field as $\eta = \psi_1$ (and correspondingly the quasi-hole's as $\zeta = \sigma_1$).
  We thus write $z_0(\tau, \{\mathcal{O}\}, \{\bar{\mathcal{O}}\}) = w_0 \eta\bar{\eta}$, with $w_0$ effectively serving as the bare electron tunneling amplitude $\Gamma_\mathrm{ET}$ in the dual regime.
  (We demonstrate in Appendix~\ref{app:inverse_duality_RR} that a toy model using the inverse duality transformation of the $\mathcal{O} = \sigma_2$ channel analytically reproduces the original $e/5$ quasi-particle tunneling process.)

  Therefore, in the dual picture, we conclude that
  \begin{align}
    S_\mathrm{dual}\qty[Θ, ψ_1, \bar{ψ}_1^\dagger]
    =& \sum_{ω}\frac{|ω|}{4πν}Θ(-ω)Θ(ω)\notag\\
     & -∫ \dd{τ} w_0 \qty[ ψ_1\bar{ψ}_1^\dagger \exp(i\frac{Θ(τ)}{ν}) + \qhc ].
    \label{eq:RR_eff_action_ET_}
  \end{align}
  This is the action for the electron tunneling process in the weak coupling regime.
  Thus, we find the duality between the quasi-particle tunneling in the strong coupling regime and the electron tunneling in the weak coupling regime even for Read--Rezayi states by considering the phase-shift instanton process as shown in \cref{fig:phase-shift_instanton}(b).

\section{Differential conductance behavior}
\label{sec:conductance}
Finally, we discuss the differential conductance behavior based on the RG analysis using the duality derived in this paper.
The detailed calculation of the RG analysis is shown in Appendix \ref{app:conductance_scaling}.
First of all, we consider the quasi-particle tunneling process.
The $β$-function for the quasi-particle tunneling amplitude $Γ_\QPT$ is given by
\begin{align}
  \dv{Γ_\QPT}{ℓ} = \qty[1-Δ_\QPT]Γ_\QPT,
  \label{eq:beta_function_QPT}
\end{align}
where $ℓ$ is the RG scale, and $Δ_\QPT$ is the scaling dimension of the quasi-particle tunneling operator.
According to \cref{eq:QPT_Hamiltonian}, the scaling dimension $Δ_\QPT$ is given by the sum of the scaling dimensions of the charge and primary fields as
\begin{align}
  Δ_\QPT = 2\qty(Δ_{α=1/n}+Δ_{ζ}),
\end{align}
where $Δ_{α}$ and $Δ_{ζ}$ are the scaling dimensions of the vertex operator $e^{iαϕ}$ and the primary field $ζ$, respectively (see Appendix~\ref{app:scaling_dimension} for details).
For each FQH state, a straightforward calculation yields the following result:
\begin{align}
  \Delta_\QPT =
  \left\{
    \begin{array}{ll c l}
      2\biggl(\Delta_{1} + \Delta_{I}\biggr)                &= \nu                              & & \text{(Laughlin)} \\
      2\biggl(\Delta_{\frac{1}{2}} + \Delta_{\sigma}\biggr) &= \dfrac{\nu}{4} + \dfrac{1}{8}    & & \text{(Moore--Read)} \\
      2\biggl(\Delta_{\frac{1}{3}} + \Delta_{\sigma_1}\biggr) &= \dfrac{\nu}{9} + \dfrac{2}{15}  & & \text{(Read--Rezayi)}
    \end{array}
  \right. .
\end{align}

Thus, the $β$-function \eqref{eq:beta_function_QPT} becomes positive for all FQH states because $Δ_\QPT < 1$, and the quasi-particle tunneling process is relevant.

When the quasi-particle tunneling amplitude $Γ_\QPT$ becomes large, however, it is not necessary to evaluate a non-perturbative analysis directly.
Using the duality derived in this paper, we can efficiently analyze the weak coupling regime of the electron tunneling process instead.
According to \cref{eq:ET_Hamiltonian}, the scaling dimension $Δ_\ET$ is also counted by the sum of the scaling dimensions of the charge and primary fields as
\begin{align}
  Δ_\ET = 2\qty(Δ_{α=1/ν}+Δ_{η}).
\end{align}
where $Δ_{η}$ is the scaling dimension of the primary field $η$, respectively.
For each FQH state, a straightforward calculation yields also the following result:
\begin{align}
  \Delta_\ET =
  \left\{
    \begin{array}{ll c l}
      2\biggl(\Delta_{\frac{1}{\nu}} + \Delta_{I}\biggr)      &= \dfrac{1}{\nu}            & & \text{(Laughlin)} \\
      2\biggl(\Delta_{\frac{1}{\nu}} + \Delta_{\psi}\biggr)   &= \dfrac{1}{\nu} + 1        & & \text{(Moore--Read)} \\
      2\biggl(\Delta_{\frac{1}{\nu}} + \Delta_{\psi_1}\biggr) &= \dfrac{1}{\nu} + \dfrac{4}{3} & & \text{(Read--Rezayi)}
    \end{array}
  \right. .
\end{align}
Therefore, as summarized in \cref{appeq:conductance_scaling} in Appendix~\ref{app:conductance_scaling}, the differential conductance $G$ is given by
\begin{align}
  G ∝
  \begin{cases}
    V^{2/ν-2}  & \text{(Laughlin)}\\
    V^{2/ν}  & \text{(Moore--Read)}\\
    V^{2/ν+2/3}  & \text{(Read--Rezayi)}
  \end{cases},
  \label{eq:conductance_scaling}
\end{align}
where $V$ is the bias voltage. 
By explicitly substituting the corresponding filling factors for the most prominent states ($\nu=1/3$ for Laughlin, $1/2$ for Moore--Read, and $3/5$ for Read--Rezayi), we find that all these seemingly distinct expressions remarkably converge to the exact same scaling:
\begin{align}
  G \propto V^4.
\end{align}
This striking universality in the strong coupling regime is physically profound.
While the weak coupling quasi-particle tunneling exhibits distinct scaling behaviors depending on the fractional charge and anyonic statistics, the strong coupling regime requires the tunneling particle across the vacuum gap to be a true physical fermion ($\Delta_\mathrm{el} = 3/2$).
Our phase-shift instanton analysis explicitly demonstrates that the complex non-Abelian fractional quasi-particles perfectly reconstruct the universal physical electron, leading to the universal $G \propto V^4$ scaling.

\section{Conclusion and Discussion} 
\label{sec:conclusion}
In this paper, we have discussed the duality between the quasi-particle tunneling process in the strong coupling regime and the electron tunneling process in the weak coupling regime for various FQH states including non-Abelian states such as the Moore--Read and Read--Rezayi states.
To derive the duality, we have employed the phase-shift instanton method for non-Abelian FQH states.
While a fully rigorous mathematical proof of the duality transformation for the Read--Rezayi state remains an open challenge, our approach establishes a physically consistent framework.
Using this duality, we can analyze the previously inaccessible strong coupling regime of the quasi-particle tunneling process by investigating the weak coupling regime of the electron tunneling process as shown in \cref{fig: MR_Geometry}(c).
This duality will be useful to understand the tunneling phenomena in various FQH states.

Additionally, in order to make a direct connection to experiment, we have evaluated the non-linear differential conductance based on the renormalization-group analysis using the duality.
Our analysis reveals a striking universality: while the weak quasi-particle tunneling distinguishes these FQH states through different power laws, their strong coupling limits universally converge to $G \propto V^4$.
Our phase-shift instanton method successfully captures how complex non-Abelian anyons, such as $\mathbb{Z}_3$ parafermions, perfectly recombine into this universal fermionic electron under the duality mapping.
Therefore, the simple two-terminal nonlinear conductance cannot distinguish these non-Abelian states in the strong coupling regime.
However, the exact dual Hamiltonian constructed in this work provides a solid foundation for evaluating more complex transport observables that may capture the residual non-Abelian statistics.

We briefly discuss directions where tunneling duality with the phase-shift instanton framework can be applied and extended.
First, it can be extended to calculate higher-order transport signatures such as shot noise and Fano factors, which have been proposed as decisive probes for the $\mathbb{Z}_3$ parafermion statistics in the Read--Rezayi state~\cite{law_2008_Probing}.
For another instance, heterostructure duality provides an application of the present analysis: the duality of the FQH states can be incorporated into microscopic point-contact Hamiltonians to study scaling of conductance with Andreev-like reflection~\cite{sandler_1998_Andreev,ohashi_2022_Andreevlike}.
Recently, experiments of the observation of the Andreev-like reflection have been reported~\cite{hashisaka_2021_Andreev,cohen_2023_Universal}, and this work will serve as a milestone for extending those experiments.

Overall, the phase-shift instanton provides a compact and physically transparent extension of tunneling duality to non-Abelian FQH states, and it opens routes toward modeling of transport and interference in engineered quantum Hall heterostructures.


\acknowledgments
R.O. is grateful to Y. Tanaka for fruitful discussions.
This work was supported by 
JSPS KAKENHI (
  Grants 
  No. JP24H00197
  No. JP24K06926,
  No. JP25H01250, 
  No. JP25K23361,
  and No. JP26K00660
).


\appendix
\begin{appendix}

\section{Mapping to Caldeira-Leggett model}
\label{app:CL_mapping}
We derive the action for the phase variable at the tunneling point.
Partion function in strong coupling regime of quasi-particle tunneling is given by
\begin{multline}
  ∫ \DD{ϕ} \exp\bigl[-S_\mathrm{U(1)+QPT}\bigr]\\
  = ∫ \DD{ϕ} \exp\biggl[
    -∫\dd{τ}\dd{x}\biggl\{\frac{1}{4πν}\biggl[\frac{1}{v}\qty(\pdv{ϕ}{τ})²+v\qty(\pdv{ϕ}{x})²\biggr]\biggr\}\\
     \hspace{20mm}  + ∫\dd{τ}Γ_\QPTζ\bar{ζ}\cos(\frac{ϕ(x=0)}{n})
  \biggr],
\end{multline}
where $ζ(z)$ is the primary operator defined in \cref{eq:zeta_eta_operator}.

We map the partition function to the Caldeira-Leggett model. 
First, we introduce an auxiliary field $θ$ which represents the phase variable at the tunneling point $x=0$:
\begin{align}
  &∫\DD{ϕ} \exp\biggl[-S_\mathrm{U(1)}+∫\dd{τ}Γ_\QPTζ\bar{ζ}\cos(\frac{ϕ(x=0,τ)}{n})\biggr]\notag\\
  =& ∫\DD{ϕ}\DD{θ} \; δ\qty(θ(τ)-ϕ(x=0,τ)) \notag\\
    & × \exp\qty[-S_\mathrm{U(1)}+Γ_\QPT∫\dd{τ}ζ\bar{ζ}\cos(\frac{θ(τ)}{n})]\notag\\
  =& ∫\DD{ϕ}\DD{θ} ∫\DD{λ} \exp\qty[∫\dd{τ}iλ(τ)\biggl(θ(τ)-ϕ(x=0,τ)\biggr)] \notag\\
    & × \exp\qty[-S_\mathrm{U(1)}+Γ_\QPT∫\dd{τ}ζ\bar{ζ}\cos(\frac{θ(τ)}{n})]\notag\\
  =& ∫\DD{θ}\DD{λ} \exp\biggl[i∫\dd{τ}λ(τ)θ(τ)\notag\\
   &\hspace{25mm}+Γ_\QPT∫\dd{τ}ζ\bar{ζ}\cos(\frac{θ(τ)}{n})\biggr]\notag\\
   &×∫\DD{ϕ}\exp\left[-S_\mathrm{U(1)}-i∫\dd{τ}λ(τ)ϕ(x=0,τ)\right]
  \label{appeq:Laughlin_Z_U1+QPT_1}.
\end{align}
Applying the Fourier transform:
\begin{equation}
  \left\{
  \begin{aligned}
    ϕ(x,τ) &= \frac{1}{\sqrt{β}}\sum_{ω}∫\frac{\dd{k}}{2π}ϕ(k,ω)e^{ikx}e^{iωτ}\\
    λ(τ)    &= \frac{1}{\sqrt{β}}\sum_{ω}λ(ω)e^{iωτ}
  \end{aligned}
  \right.,
  \label{appeq:Fourier_transform}
\end{equation}
the partition function for $ϕ$ field is written as
\begin{align}
  &∫\DD{ϕ}\exp\left[-S_\mathrm{U(1)}-i∫\dd{τ}λ(τ)ϕ(x=0,τ)\right]\notag\\
  =&∫\DD{ϕ} \exp\Biggl[\sum_{ω}\biggl(-iλ(ω)∫\frac{\dd{k}}{2π}ϕ(k,ω)\notag\\
    &\hspace{7mm}-∫ \frac{\dd{k}}{2π}ϕ(-k,-ω)\qty[\frac{1}{8πν}\qty(\frac{ω²}{v}+vk²)]ϕ(k,ω)\biggr)\Biggr]\notag\\
  =&∫\DD{ϕ} \exp\Biggl[\sum_{ω}\biggl(∫ \frac{\dd{k}}{2π}\biggl\{
    -iλ(ω)ϕ(k,ω)\notag\\
    &\hspace{10mm}-\frac{1}{4}ϕ(-k,-ω)\qty[\frac{1}{2πν}\qty(\frac{ω²}{v}+vk²)]ϕ(k,ω)\biggr\}\biggr)\Biggr].
    \label{appeq:Laughlin_Z_phi_int}
\end{align}
Integrating out $ϕ$ with Gaussian integration formula
\begin{align}
  \text{\eqref{appeq:Laughlin_Z_phi_int}}
  =&\exp\biggl[\sum_{ω}\biggl(-∫\frac{\dd{k}}{2π}
    λ(-ω)\biggl[\frac{2πν}{\qty(ω²/v+vk²)}\biggr]λ(ω)\biggr)\biggr]\notag\\
  =&\exp\biggl[\sum_{ω}\qty(-λ(-ω)λ(ω)\frac{πν}{|ω|})\biggr].
  \label{appeq:Laughlin_Z_phi}
\end{align}
Then, integrating out $λ$
\begin{align}
  &\eqref{appeq:Laughlin_Z_U1+QPT_1}\notag\\
  =& ∫\DD{θ}\exp\biggl[+Γ_\QPT∫\dd{τ}ζ\bar{ζ}\cos(\frac{θ(τ)}{n})\biggr]\notag\\
   &×∫\DD{λ} \exp\biggl[-\sum_{ω}λ(-ω)λ(ω)\frac{πν}{|ω|}+i∫\dd{τ}λ(τ)θ(τ)\biggr]\notag\\
  =& ∫\DD{θ}\exp\Biggl[-\sum_{ω}\frac{|ω|}{4πν}θ(ω)θ(-ω)\notag\\
  &\hspace{25mm}+Γ_\QPT∫\dd{τ}ζ\bar{ζ}\cos(\frac{θ(τ)}{n})\Biggr].
  \label{appeq:Laughlin_Z_U1+QPT_2}
\end{align}
Thus the action for the phase variable $θ(τ)$ is given by
\begin{multline}
  S_\mathrm{U(1)+QPT}
  = \sum_{ω}\frac{|ω|}{4πν}θ(-ω)θ(ω)\\
    -∫ \dd{τ}Γ_\QPTζ\bar{ζ}\cos(\frac{θ(τ)}{n}).
    \label{appeq:Laughlin_eff_action}
\end{multline}
In the view point of the dynamics for $θ$, the first term in \cref{appeq:Laughlin_eff_action} is the friction term in the Caldeira-Leggett theory, and the second term is regarded as the periodic potential.

\section{Stratonovich--Hubbard transformation}
\label{app:SH_transformation}
We derive the dual action for the strong coupling regime of the (phase-shifted) instanton partition function in \cref{eq:instanton_gas_partition_function} using the Stratonovich--Hubbard transformation.
The instanton partition function is given by:
\begin{align}
  Z_\mathrm{ins}
    =& \sum_{n=0}^{∞}\sum_{\qty{eᵢ}}
    \frac{1}{n!}
    \prod_{ℓ=0}^{n} ∫_{0}^{β}\dd{τ_{ℓ}}z₀(τ_{\ell})\notag\\
    &×\exp\qty[
      -\,\frac{1}{\nu^2}\sum_{ij}^{n}
      e_i e_j
      J(τᵢ - τⱼ)
    ]
    \label{appeq:Z_U1+QPT}
    ,\\
    \because\quad
    J(τ) =& \frac{πν}{β}\sum_{ω}\frac{e^{-iωτ}}{|ω|}\qc{} J(ω) = \frac{πν}{\sqrt{β}|ω|}.
    \label{appeq:J_kernel}
\end{align}
Introducing the Stratonovich--Hubbard field $Θ(τ)$, we rewrite the interaction term
\begin{multline}
  \exp\qty[
    -\frac{1}{ν^2}\sum_{ij}^{n} e_i e_j J(τ_i - τ_j)
  ]\\
  = ∫ \DD{Θ}\,
  \exp\Biggl[
    -\frac{1}{4}∫ \dd{τ}\dd{τ'} Θ(τ)J^{-1}(τ - τ')Θ(τ')\\
    + \frac{i}{ν}\sum_j e_j Θ(τ_j)
  \Biggr],
  \label{appeq:SH_transformation}
\end{multline}
where $J^{-1}(τ)$ is the inverse kernel.
Substituting and performing Fourier transformation:
\begin{align}
  &∫\dd{τ}\dd{τ'} Θ(τ)J^{-1}(τ-τ')Θ(τ')\notag\\
  &= \frac{1}{\sqrt{β³}}\sum_{ω,ω₁,ω₂}Θ(ω₁)J^{-1}(ω)Θ(ω₂)\notag\\
    &\hspace{20mm}×∫\dd{τ}\dd{τ'}e^{i(ω₁+ω)τ}e^{i(ω₂-ω)τ'}\notag\\
  &= \frac{1}{\sqrt{β³}}\sum_{ω,ω₁,ω₂}
     Θ(ω₁)J^{-1}(ω)Θ(ω₂)\,
     βδ_{ω₁+ω,0}\,βδ_{ω₂-ω,0}\notag\\
  &= \frac{1}{\sqrt{β}}\sum_{ω} J^{-1}(ω)Θ(-ω)Θ(ω).
  \label{appeq:inverse_kernel_Fourier}
\end{align}
Thus the Gaussian part becomes
\begin{align}
  &\exp\qty[
    -\frac{1}{4}∫ \dd{τ}\dd{τ'}
    Θ(τ)J^{-1}(τ-τ')Θ(τ')
  ]\notag\\
  &\hspace{20mm}= 
  \exp\qty[
    -\frac{1}{4\sqrt{β}}\sum_{ω} J^{-1}(ω)\,Θ(-ω)Θ(ω)
  ].
  \label{appeq:Gaussian_part}
\end{align}
Using \cref{appeq:J_kernel}, we obtain
\begin{multline}
  Z_\mathrm{ins} 
  = ∫ \DD{Θ}\sum_{n=0}^{∞}\sum_{\qty{eᵢ}}
   \frac{1}{n!}
   \prod_{ℓ=0}^{n} ∫_{0}^{β}\dd{τ_{ℓ}}z₀(τ_{\ell})\\
   ×\exp\qty[
    -\sum_{ω} \frac{|ω|}{4πν}\,Θ(-ω)Θ(ω) + \frac{i}{ν}\sum_{j} e_j Θ(τ_j)
   ].
   \label{appeq:Z_U1+QPT_SH}
\end{multline}
We set $z_0(\tau)$ is a fugacity of the instanton.
Now, we perform the summation over $eᵢ=±1$ in \cref{appeq:Z_U1+QPT_SH}:
\begin{align}
  &Z_\mathrm{ins} \notag\\
  =& ∫ \DD{Θ}
   \exp\qty[
     -\sum_{ω}\frac{|ω|}{4πν}Θ(-ω)Θ(ω)
   ]\notag\\
  &×\sum_{n=0}^{∞}\sum_{\qty{e_i}} \frac{1}{n!}
  \prod_{ℓ=0}^{n}\dd{τ_ℓ}∫_{0}^{β}
    z₀(τ_ℓ)
   \,
   \exp\qty[
      i\sum_i e_i \frac{Θ(τ_i)}{ν}
   ]\notag\\
  =& ∫ \DD{Θ}
   \exp\qty[
     -\sum_{ω}\frac{|ω|}{4πν}Θ(-ω)Θ(ω)
   ]\notag\\
  &\hspace{1mm}×\sum_{n=0}^{∞} \frac{1}{n!}
  \prod_{ℓ=0}^{n}\dd{τ_ℓ}∫_{0}^{β}
   z₀(τ_ℓ)
   \qty(e^{i\frac{Θ(τ_i)}{ν}}+e^{-i\frac{Θ(τ_i)}{ν}})
   \notag\\
  =& ∫ \DD{Θ}
   \exp\qty[
     -\sum_{ω}\frac{|ω|}{4πν}Θ(-ω)Θ(ω)
   ]\notag\\
  &\hspace{1mm}×\sum_{n=0}^{∞} \frac{1}{n!}
  \prod_{ℓ=0}^{n}\dd{τ_ℓ}∫_{0}^{β}
   2z₀(τ_ℓ)\cos\qty(\frac{Θ(τ_ℓ)}{ν})
   \notag\\
  =& ∫ \DD{Θ}
   \exp\qty[
     -\sum_{ω}\frac{|ω|}{4πν}Θ(-ω)Θ(ω)
   ]\notag\\
  &\hspace{1mm}×\sum_{n=0}^{∞} \frac{1}{n!}
    \qty[∫_{0}^{β}\dd{τ}2z₀(τ)\cos\qty(\frac{Θ(τ)}{ν})]^{n}\notag\\
  =& ∫ \DD{Θ}
   \exp\qty[
     -\sum_{ω}\frac{|ω|}{4πν}Θ(-ω)Θ(ω)
   ]\notag\\
   &\hspace{10mm} ×\exp\qty[∫_{0}^{β}\dd{τ}2z₀(τ)\cos\qty(\frac{Θ(τ)}{ν})].
   \label{appeq:Z_U1+QPT_final}
\end{align}
We thus obtain the dual action for $Θ$ field:
\begin{align}
  S_\mathrm{dual}
  =& \sum_{ω}\frac{|ω|}{4πν}Θ(-ω)Θ(ω)
    -∫ \dd{τ}2z₀\cos\qty(\frac{Θ(τ)}{ν}).
    \label{appeq:Laughlin_eff_action_Theta}
\end{align}
Note that this transformation holds for the Laughlin states.
For the Moore--Read and Read--Rezayi states, since the expectation values of the CFT fields are required, the instanton fugacity $z_0$ generally takes the form $z_0(\tau, \{\mathcal{O}\}, \{\bar{\mathcal{O}}\})$.
As discussed in the main text, the coefficient of $\Theta$ in the exponential term indicates a tunneling process of a particle with charge $e$.
Therefore, the conformal family $\{\mathcal{O}\}$ in the fugacity is restricted to the candidates carrying charge $e$.
Although this mathematical mapping alone cannot strictly restrict $\mathcal{O}$ to a unique primary field, physical requirements strongly select the true electron operator's primary field $\eta$ as the most plausible candidate.
We thus write $z_0(\tau, \{\mathcal{O}\}, \{\bar{\mathcal{O}}\}) = w_0 \eta\bar{\eta}$, where the coefficient $w_0$ represents the component independent of the conformal family, effectively serving as the bare electron tunneling amplitude $\Gamma_\mathrm{ET}$ in the dual regime.

\section{Analysis of the differential conductance}
\label{app:conductance_scaling}
We review the scaling behavior on the point contact system using the renormalization-group (RG) method.\cite{fradkin_2013_Field}
Let us consider a system close to a fixed point.
We can write the action of the tunneling system in the form
\begin{align}
  S(\phi) = S^{\ast}(\phi) - ∫\dd{x}\dd{τ}\; Γδ(x)\; \mathcal{T}(ϕ(x,τ)),
\end{align}
where $Γ$ is a tunneling coupling constant, the tunneling operator $\mathcal{T}(x)$ has the scaling dimension $Δ$, and $S^{\ast}$ is fixed point action.
Now, we consider the point contact system, thus, we introduce the delta function $δ(x)$ in the tunneling term, and we can integrate out the spacial coordinate $x$.
Then, the action is rewritten as
\begin{align}
  S(\phi) = S^{\ast}(\phi) - ∫\dd{τ}\; Γ\; \mathcal{T}(ϕ(τ)).
\end{align}
Under a RG transformation consisting of integrating out high-energy modes $Λ → bΛ$ ($b<1$), which is equal to $τ → τ/b$, and the action $S^{\ast}(\phi)$ remains invariant since it is a fixed point.
The operator $\mathcal{T}(ϕ(τ))$ transforms as
\begin{align}
 \mathcal{T}(ϕ(b^{-1}τ)) = b^{Δ}\mathcal{T}(ϕ(τ)).
\end{align}
Then, under the action of the RG the perturbation transoforms as
\begin{align}
  ∫\dd{τ}Γ \mathcal{T}(ϕ(τ)) → ∫\dd{τ}Γ b^{-1+Δ}\mathcal{T}(ϕ(τ)).
\end{align}
Thus, the RG transformation is equivalent to a rescaling of the coupling constant
\begin{align}
  Γ^{\prime} = Γ b^{-1+Δ} ≡ Γ(b).
\end{align}
We can write $b=e^{-ℓ}=Λ^{\prime}/Λ₀$ where $Λ₀$ and $Λ^{\prime}$ are bare and renormalized cutoff, respectively.
And turn the transformation into a differential change of the coupling constant of term of the $β$-function
\begin{align}
  \dv{Γ}{ℓ} = (1-Δ)Γ
  \label{appeq:beta_function_1}
\end{align}
or in terms of the cut-off $Λ^{\prime}$
\begin{align}
  Λ^{\prime}\dv{Γ}{Λ^{\prime}} = -(1-Δ)Γ.
  \label{appeq:beta_function_2}
\end{align}
By intergrating the $β$-function \eqref{appeq:beta_function_2}, we obtain the scaling behavior of the coupling constant
\begin{align}
  \qty(\frac{Γ(Λ^{\prime})}{Γ(Λ₀)}) = \qty(\frac{Λ^{\prime}}{Λ₀})^{-(1-Δ)}.
\end{align}
The current $I$ is proportional to $|Γ(Λ^{\prime})|²V$, thus, we obtain the $I$-$V$ characteristics derives as follow
\begin{align}
  I ∝ \qty(\frac{Γ(Λ₀)}{Λ₀^{-(1-Δ)}})² \qty(Λ^{\prime})^{-2(1-Δ)} V.
\end{align}
Finally, at high-bias $eV>k_\mathrm{B}T$, the renormalized cut-off $Λ^{\prime}$ can be $V$:
\begin{align}
  I ∝ V^{2Δ - 1}
\end{align}
and the differential conductance $G=\dd{I}/\dd{V}$ is also given by
\begin{align}
  G ∝ V^{2(Δ-1)}.
  \label{appeq:conductance_scaling}
\end{align}
Therefore, we can obtain the scaling behavior of the differential conductance to count the scaling dimension of the tunneling operator.

\section{Scaling dimension of primary fields}
\label{app:scaling_dimension}
Here, we summarize the scaling dimension of the primary fields used in this paper.\cite{francesco_2012_Conformal}
Firstly, we discuss the vertex operator $V_{α}$ defined as
\begin{align}
  V_{α} ≡ e^{iαϕ}
\end{align}
which constructs the free boson $ϕ$ which obeys a following action:
\begin{align}
  S_\mathrm{boson} = \frac{1}{2}g∫\dd{τ}\dd{x}\; \qty[\qty(\pdv{ϕ}{τ})² + \qty(\pdv{ϕ}{x})²],
\end{align}
where $g$ is some normalization constant depending on the system.
We can obtain the scaling dimension of the vertex operators by calculating the operator expansion between the energy-momentum tensor and $V_{α}$.
The detail calculation is shown in Ref.~\citenum{francesco_2012_Conformal}, and here we only summarize the result as follows:
\begin{align}
  Δ_{α} = \frac{α²}{8πg}.
  \label{appeq:scaling_dimension_vertex_general}
\end{align}
In the our system, the action for the $U(1)$ charge sector is given by \cref{eq:boson_action_even_odd,eq:S_U1}, thus, we can chose the normalization constant as $g=1/4πν$.
Therefore, the scaling dimension of the vertex operator in our system is as follows:
\begin{align}
  Δ_{α} = \frac{ν}{2}α².
  \label{appeq:scaling_dimension_vertex}
\end{align}
Next, we summarize the scaling dimension of the primary field in the Ising CFT as follows\cite{francesco_2012_Conformal}:
\begin{align}
  Δ_{I}=0\qc{} Δ_{ψ} = \frac{1}{2}\qc{} Δ_{σ} = \frac{1}{16}.
  \label{appeq:scaling_dimension_Ising}
\end{align}
Finally, we show the scaling dimension of some primary fields in the $\mathbb{Z}_{3}$ parafermion CFT.
In this paper, we use only parafermion field $ψ₁$ and spin field $σ₁$, whose scaling dimensions are given by\cite{francesco_2012_Conformal}
\begin{align}
  Δ_{ψ_i} = \frac{2}{3}\qc{} Δ_{σ_i} = \frac{1}{15},
  \label{appeq:scaling_dimension_Z3}
\end{align}
where $i=1,2$, and the 
\begin{align}
  Δ_{ε} = \frac{1}{5}.
  \label{appeq:scaling_dimension_Z3_2}
\end{align}

\section{Inverse duality transformation of the \texorpdfstring{$k=3$}{k=3} Read--Rezayi state}
\label{app:inverse_duality_RR}

In the main text, we demonstrated the duality transformation from the weak coupling quasi-particle tunneling regime to the strong coupling electron tunneling regime using the phase-shift instanton method.
To confirm the mathematical consistency of our framework, we investigate the inverse duality transformation---from the strong coupling electron tunneling regime back to the weak coupling quasi-particle tunneling regime---for the Read--Rezayi state. 

As discussed in \cref{sec:duality_RR}, the candidates for the electron tunneling operator (carrying charge $e$) in the Read-–Rezayi state are $\mathcal{O}=\psi_1$ and $\mathcal{O}=\sigma_2$.
While the non-local parafermion field $\psi_1$ is difficult to evaluate directly within the instanton-gas approximation, the $\sigma_2$ field belongs to the spin operator family of the three-state Potts model and possesses a well-defined matrix representation.
Therefore, we consider a toy model of electron tunneling mediated by the $\sigma_2$ channel to perform the inverse duality transformation analytically.

Let us start with the dual action for the electron tunneling process mediated by $\eta = \sigma_2$:
\begin{align}
  S_{\mathrm{ET}}[\Theta, \sigma_2, \bar{\sigma}_2] =& \sum_{\omega} \frac{|\omega|}{4\pi\nu} \Theta(-\omega)\Theta(\omega) \notag\\
      &-∫ \dd{τ}\frac{Γ_\mathrm{ET}}{2}\qty( σ_2\bar{σ}_2^{\dagger} e^{i\Theta(τ)/\nu}+\qhc).
\end{align}
The field $\sigma_2$ is the Hermitian conjugate of $\sigma_1$, $\hat{\sigma}_2 = \hat{\sigma}_1^\dagger$, and its matrix representation is given by
\begin{align}
  \hat{\sigma}_2 = \begin{pmatrix}
  1 & 0 & 0 \\
  0 & e^{-i 2\pi/3} & 0 \\
  0 & 0 & e^{-i 4\pi/3}
  \end{pmatrix}.
\end{align}
Because of the internal degrees of freedom of $\sigma_2$, the periodic potential $\cos(\Theta/\nu)$ mathematically experiences a minimum phase shift of $\Delta(\Theta/\nu) = 2\pi/3$.
The instantons will automatically select the path of least resistance (i.e., the shortest path) to minimize the instanton action.
Thus, the corresponding step size of the $\Theta$ field for the phase-shift instanton is exactly determined as $\Delta\Theta = h_0 = 2\pi\nu/3$.

Following the instanton-gas approximation procedure discussed in \cref{eq:Laughlin_S_ins}, the instanton interaction for this step size becomes
\begin{align}
  S_{\mathrm{ins}}[\Theta] \propto \frac{h_0^2}{4\pi\nu} = \frac{(2\pi\nu/3)^2}{4\pi\nu} = \frac{\pi\nu}{9}.
\end{align}
By applying the Stratonovich--Hubbard transformation to this instanton gas, we introduce the original phase field $\phi$ governing the weak coupling regime.
Since the interaction coefficient is exactly $\pi\nu/9$, the transformation uniquely generates a weak coupling tunneling potential of the form $\exp(i\phi/n^{\ast})+\qhc$, where the coefficient satisfies
\begin{align}
  \frac{\pi\nu}{(n^*)^2} = \frac{\pi\nu}{9} \implies n^* = 3.
\end{align}
This result perfectly reproduces the quasi-particle tunneling operator of the Read--Rezayi state with fractional charge $e/5$ ($n=3$)
\begin{align}
  S_\mathrm{inv-dual}[\theta,\zeta,\bar{\zeta}]
  =& \sum_{ω}\frac{|ω|}{4πν}\theta(-ω)\theta(ω)\notag\\
    &\hspace{2mm} -∫ \dd{τ}z₀\qty(\zeta\bar{\zeta}^{\dagger} e^{i\theta(\tau)/3}+ \qhc)
  \label{eq:inv_RR_eff_action_QPT}
\end{align}
and the primary field $\zeta=\sigma_1$ or $\psi_2$.

Finally, we consider the scaling dimension of this model.
If the $\sigma_2$ channel could physically mediate the electron tunneling, the scaling dimension $\Delta_{\mathrm{ET}, \sigma_2}$ would be given by the sum of the charge sector and the $\sigma_2$ primary field:
\begin{align}
  \Delta_{\mathrm{ET}, \sigma_2} = 2 \left( \Delta_{1/\nu} + \Delta_{\sigma_2} \right) = 2 \left( \frac{5}{6} + \frac{1}{15} \right) = \frac{9}{5}.
\end{align}
According to \cref{appeq:conductance_scaling}, the corresponding differential conductance would scale as
\begin{align}
  G \propto V^{2(\Delta_{\mathrm{ET}, \sigma_2}-1)} = V^{8/5}.
\end{align}
As discussed in the main text, since $\Delta_{\mathrm{ET}, \sigma_2} = 9/5 \neq 3/2$, the operator $\sigma_2 e^{i\phi/\nu}$ is an anyon and cannot physically tunnel across the vacuum gap.
The true physical electron tunneling is exclusively mediated by the fermion channel $\psi_1$ yielding the universal $G \propto V^4$ scaling.
Nevertheless, the $\sigma_2$ toy model mathematically illuminates the robustness of the duality mapping.

\end{appendix}

\bibliographystyle{jpsj} 
\bibliography{reference.bib} 

@article{laughlin_1983_Anomalous,
  title = {Anomalous {{Quantum Hall Effect}}: {{An Incompressible Quantum Fluid}} with {{Fractionally Charged Excitations}}},
  shorttitle = {Anomalous {{Quantum Hall Effect}}},
  author = {Laughlin, R. B.},
  year = {1983},
  date = {1983-05-02},
  journal = {Phys. Rev. Lett.},
  volume = {50},
  number = {18},
  pages = {1395--1398},
  publisher = {American Physical Society},
  doi = {10.1103/PhysRevLett.50.1395}
}

@article{tsui_1982_TwoDimensional,
  title = {Two-{{Dimensional Magnetotransport}} in the {{Extreme Quantum Limit}}},
  author = {Tsui, D. C. and Stormer, H. L. and Gossard, A. C.},
  year = {1982},
  date = {1982-05-31},
  journal = {Phys. Rev. Lett.},
  volume = {48},
  number = {22},
  pages = {1559--1562},
  publisher = {American Physical Society},
  doi = {10.1103/PhysRevLett.48.1559}
}

@article{klitzing_1980_New,
  title = {New Method for High-Accuracy Determination of the Fine-Structure Constant Based on Quantized Hall Resistance},
  author = {Klitzing, K. v. and Dorda, G. and Pepper, M.},
  year = {1980},
  journal = {Phys. Rev. Lett.},
  volume = {45},
  issue = {6},
  pages = {494--497},
  numpages = {0},
  year = {1980},
  month = {Aug},
  publisher = {American Physical Society},
  doi = {10.1103/PhysRevLett.45.494},
  url = {https://link.aps.org/doi/10.1103/PhysRevLett.45.494}
}

@book{prange_1990_Quantum,
  title = {The {{Quantum Hall Effect}}},
  editor = {Prange, Richard E. and Girvin, Steven M.},
  editora = {Birman, Joseph L. and Faissner, H. and Lynn, Jeffrey W.},
  editoratype = {redactor},
  date = {1990},
  series = {Graduate {{Texts}} in {{Contemporary Physics}}},
  publisher = {Springer New York},
  location = {New York, NY},
  doi = {10.1007/978-1-4612-3350-3},
  isbn = {978-0-387-97177-3 978-1-4612-3350-3}
}

@article{kivelson_1985_Consequences,
  title = {Consequences of Gauge Invariance for Fractionally Charged Quasi-Particles},
  author = {Kivelson, S. and Ro\v{c}ek, M.},
  year = {1985},
  date = {1985-06-13},
  journal = {Phys. Lett. B},
  volume = {156},
  number = {1},
  pages = {85--88},
  issn = {0370-2693},
  doi = {10.1016/0370-2693(85)91359-0}
}

@article{arovas_1984_Fractional,
  title = {Fractional {{Statistics}} and the {{Quantum Hall Effect}}},
  author = {Arovas, Daniel and Schrieffer, J. R. and Wilczek, Frank},
  year = {1984},
  date = {1984-08-13},
  journal = {Phys. Rev. Lett.},
  volume = {53},
  number = {7},
  pages = {722--723},
  publisher = {American Physical Society},
  doi = {10.1103/PhysRevLett.53.722}
}

@article{jain_1989_Compositefermion,
  title = {Composite-Fermion Approach for the Fractional Quantum {{Hall}} Effect},
  author = {Jain, J. K.},
  year = {1989},
  date = {1989},
  journal = {Phys. Rev. Lett.},
  volume = {63},
  number = {2},
  pages = {199--202},
  doi = {10.1103/PhysRevLett.63.199}
}

@article{jain_1989_Incompressible,
  title = {Incompressible Quantum {{Hall}} States},
  author = {Jain, J. K.},
  year = {1989},
  date = {1989-10-15},
  journal = {Phys. Rev. B},
  volume = {40},
  number = {11},
  pages = {8079--8082},
  publisher = {American Physical Society},
  doi = {10.1103/PhysRevB.40.8079}
}

@article{jain_1990_Theory,
  title = {Theory of the Fractional Quantum {{Hall}} Effect},
  author = {Jain, J. K.},
  year = {1990},
  date = {1990-04-15},
  journal = {Phys. Rev. B},
  volume = {41},
  number = {11},
  pages = {7653--7665},
  publisher = {American Physical Society},
  doi = {10.1103/PhysRevB.41.7653}
}

@article{willett_1987_Observation,
  title = {Observation of an Even-Denominator Quantum Number in the Fractional Quantum {{Hall}} Effect},
  author = {Willett, R. and Eisenstein, J. P. and St\"ormer, H. L. and Tsui, D. C. and Gossard, A. C. and English, J. H.},
  year = {1987},
  date = {1987-10-12},
  journal = {Phys. Rev. Lett.},
  volume = {59},
  number = {15},
  pages = {1776--1779},
  publisher = {American Physical Society},
  doi = {10.1103/PhysRevLett.59.1776}
}

@article{haldane_1985_Periodic,
  title = {Periodic {{Laughlin-Jastrow}} Wave Functions for the Fractional Quantized {{Hall}} Effect},
  author = {Haldane, F. D. M. and Rezayi, E. H.},
  year = {1985},
  date = {1985-02-15},
  journal = {Phys. Rev. B},
  volume = {31},
  number = {4},
  pages = {2529--2531},
  publisher = {American Physical Society},
  doi = {10.1103/PhysRevB.31.2529}
}

@article{wen_1995_Topological,
  title = {Topological Orders and Edge Excitations in Fractional Quantum {{Hall}} States},
  author = {Wen, Xiao-Gang},
  year = {1995},
  date = {1995-10-01},
  journal = {Advances in Physics},
  volume = {44},
  number = {5},
  pages = {405--473},
  publisher = {Taylor \& Francis},
  issn = {0001-8732},
  doi = {10.1080/00018739500101566}
}

@article{witten_1989_Quantum,
  title = {Quantum Field Theory and the {{Jones}} Polynomial},
  author = {Witten, Edward},
  year = {1989},
  date = {1989-09-01},
  journal = {Commun.Math. Phys.},
  volume = {121},
  number = {3},
  pages = {351--399},
  issn = {1432-0916},
  doi = {10.1007/BF01217730}
}

@article{witten_1992_Holomorphic,
  title = {On Holomorphic Factorization of {{WZW}} and Coset Models},
  author = {Witten, Edward},
  year = {1992},
  date = {1992-01},
  journal = {Commun.Math. Phys.},
  volume = {144},
  number = {1},
  pages = {189--212},
  publisher = {Springer},
  issn = {0010-3616, 1432-0916},
}

@article{moore_1991_Nonabelions,
  title = {Nonabelions in the Fractional Quantum Hall Effect},
  author = {Moore, Gregory and Read, Nicholas},
  year = {1991},
  date = {1991-08-19},
  journal = {Nucl. Phys. B},
  volume = {360},
  number = {2},
  pages = {362--396},
  issn = {0550-3213},
  doi = {10.1016/0550-3213(91)90407-O}
}

@article{moore_1989_Classical,
  title = {Classical and Quantum Conformal Field Theory},
  author = {Moore, Gregory and Seiberg, Nathan},
  year = {1989},
  date = {1989-06-01},
  journal = {Commun. Math. Phys.},
  volume = {123},
  number = {2},
  pages = {177--254},
  issn = {1432-0916},
  doi = {10.1007/BF01238857}
}

@article{blok_1990_Effective,
  title = {Effective Theories of the Fractional Quantum {{Hall}} Effect at Generic Filling Fractions},
  author = {Blok, B. and Wen, X. G.},
  year = {1990},
  date = {1990-11-01},
  journal = {Phys. Rev. B},
  volume = {42},
  number = {13},
  pages = {8133--8144},
  publisher = {American Physical Society},
  doi = {10.1103/PhysRevB.42.8133}
}

@article{kitaev_2003_Faulttolerant,
  title = {Fault-Tolerant Quantum Computation by Anyons},
  author = {Kitaev, A.Yu.},
  year = {2003},
  date = {2003-01},
  journal = {Ann. Phys.},
  volume = {303},
  number = {1},
  pages = {2--30},
  issn = {00034916},
  doi = {10.1016/S0003-4916(02)00018-0}
}

@book{francesco_2012_Conformal,
  title = {Conformal Field Theory},
  author = {Francesco, Philippe and Mathieu, Pierre and S\'en\'echal, David},
  year = {2012},
  publisher = {Springer Science \& Business Media}
}

@article{greiter_1991_Paired,
  title = {Paired {{Hall}} State at Half Filling},
  author = {Greiter, Martin and Wen, Xiao-Gang and Wilczek, Frank},
  year = {1991},
  date = {1991-06-17},
  journal = {Phys. Rev. Lett.},
  volume = {66},
  number = {24},
  pages = {3205--3208},
  publisher = {American Physical Society},
  doi = {10.1103/PhysRevLett.66.3205}
}

@article{morf_1998_Transition,
  title = {Transition from {{Quantum Hall}} to {{Compressible States}} in the {{Second Landau Level}}: {{New Light}} on the \$\textbackslash ensuremath\{\textbackslash nu\}\textbackslash phantom\{\textbackslash rule\{0ex\}\{0ex\}\}=\textbackslash phantom\{\textbackslash rule\{0ex\}\{0ex\}\}5/2\$ {{Enigma}}},
  shorttitle = {Transition from {{Quantum Hall}} to {{Compressible States}} in the {{Second Landau Level}}},
  author = {Morf, R. H.},
  year = {1998},
  date = {1998-02-16},
  journal = {Phys. Rev. Lett.},
  volume = {80},
  number = {7},
  pages = {1505--1508},
  publisher = {American Physical Society},
  doi = {10.1103/PhysRevLett.80.1505}
}

@article{read_2000_Paired,
  title = {Paired States of Fermions in Two Dimensions with Breaking of Parity and Time-Reversal Symmetries and the Fractional Quantum {{Hall}} Effect},
  author = {Read, N. and Green, Dmitry},
  year = {2000},
  date = {2000-04-15},
  journal = {Phys. Rev. B},
  volume = {61},
  number = {15},
  pages = {10267--10297},
  publisher = {American Physical Society},
  doi = {10.1103/PhysRevB.61.10267}
}

@article{read_1999_Paired,
  title = {Beyond Paired Quantum {{Hall}} States: {{Parafermions}} and Incompressible States in the First Excited {{Landau}} Level},
  shorttitle = {Beyond Paired Quantum {{Hall}} States},
  author = {Read, N. and Rezayi, E.},
  year = {1999},
  date = {1999-03-15},
  journal = {Phys. Rev. B},
  volume = {59},
  number = {12},
  pages = {8084--8092},
  publisher = {American Physical Society},
  doi = {10.1103/PhysRevB.59.8084}
}

@article{dotsenko_1984_Critical,
title = {Critical behaviour and associated conformal algebra of the {{Z$_3$}} Potts model},
journal = {Nuc. Phys. B},
volume = {235},
number = {1},
pages = {54-74},
year = {1984},
issn = {0550-3213},
doi = {https://doi.org/10.1016/0550-3213(84)90148-2},
url = {https://www.sciencedirect.com/science/article/pii/0550321384901482},
author = {Vl.S. Dotsenko},
}

@article{nayak_1996_2nquasihole,
  title = {2n-Quasihole States Realize 2n-1-Dimensional Spinor Braiding Statistics in Paired Quantum {{Hall}} States},
  author = {Nayak, Chetan and Wilczek, Frank},
  year = {1996},
  date = {1996-11-18},
  journal = {Nucl. Phys. B},
  volume = {479},
  number = {3},
  pages = {529--553},
  issn = {0550-3213},
  doi = {10.1016/0550-3213(96)00430-0}
}

@article{freedman_2002_Modular,
  title = {A {{Modular Functor Which}} Is {{Universal}}\P for {{Quantum Computation}}},
  author = {Freedman, Michael H. and Larsen, Michael and Wang, Zhenghan},
  year = {2002},
  date = {2002-06-01},
  journal = {Commun. Math. Phys.},
  volume = {227},
  number = {3},
  pages = {605--622},
  issn = {1432-0916},
  doi = {10.1007/s002200200645}
}

@article{nayak_2008_NonAbelian,
  title = {Non-{{Abelian}} Anyons and Topological Quantum Computation},
  author = {Nayak, Chetan and Simon, Steven H. and Stern, Ady and Freedman, Michael and Das Sarma, Sankar},
  year = {2008},
  date = {2008-09-12},
  journal = {Rev. Mod. Phys.},
  volume = {80},
  number = {3},
  pages = {1083--1159},
  publisher = {American Physical Society},
  doi = {10.1103/RevModPhys.80.1083}
}

@article{stone_1991_Vertex,
  title = {Vertex Operators in the Quantum Hall Effect},
  author = {Stone, Michael},
  year = {1991},
  date = {1991-02-10},
  journal = {Int. J. Mod. Phys. B},
  volume = {05},
  number = {03},
  pages = {509--527},
  publisher = {World Scientific Publishing Co.},
  issn = {0217-9792},
  doi = {10.1142/S0217979291000316}
}

@article{chang_2003_Chiral,
  title = {Chiral {{Luttinger}} Liquids at the Fractional Quantum {{Hall}} Edge},
  author = {Chang, A. M.},
  year = {2003},
  date = {2003-11-13},
  journal = {Rev. Mod. Phys.},
  volume = {75},
  number = {4},
  pages = {1449--1505},
  publisher = {American Physical Society},
  doi = {10.1103/RevModPhys.75.1449}
}

@book{fradkin_2013_Field,
  title = {Field Theories of Condensed Matter Physics},
  author = {Fradkin, Eduardo},
  year = {2013},
  date = {2013},
  publisher = {Cambridge University Press}
}

@article{wen_1990_Chiral,
  title = {Chiral {{Luttinger}} Liquid and the Edge Excitations in the Fractional Quantum {{Hall}} States},
  author = {Wen, Xiao-Gang},
  year = {1990},
  date = {1990-06-15},
  journal = {Phys. Rev. B},
  volume = {41},
  number = {18},
  pages = {12838--12844},
  issn = {0163-1829, 1095-3795},
  doi = {10.1103/PhysRevB.41.12838}
}

@article{wen_1991_EDGE,
  title = {{{EDGE EXCITATIONS IN THE FRACTIONAL QUANTUM HALL STATES AT GENERAL FILLING FRACTIONS}}},
  author = {Wen, X.G.},
  year = {1991},
  date = {1991-01-10},
  journal = {Mod. Phys. Lett. B},
  volume = {05},
  number = {01},
  pages = {39--46},
  issn = {0217-9849, 1793-6640},
  doi = {10.1142/S0217984991000058}
}

@article{elitzur_1989_Remarks,
  title = {Remarks on the Canonical Quantization of the {{Chern-Simons-Witten}} Theory},
  author = {Elitzur, Shmuel and Moore, Gregory and Schwimmer, Adam and Seiberg, Nathan},
  year = {1989},
  date = {1989-10-30},
  journal = {Nucl. Phys. B},
  volume = {326},
  number = {1},
  pages = {108--134},
  issn = {0550-3213},
  doi = {10.1016/0550-3213(89)90436-7}
}

@article{milovanovic_1996_Edge,
  title = {Edge Excitations of Paired Fractional Quantum {{Hall}} States},
  author = {Milovanovi\'c, M. and Read, N.},
  year = {1996},
  date = {1996-05-15},
  journal = {Phys. Rev. B},
  volume = {53},
  number = {20},
  pages = {13559--13582},
  publisher = {American Physical Society},
  doi = {10.1103/PhysRevB.53.13559}
}

@article{milliken_1996_Indications,
  title = {Indications of a {{Luttinger}} Liquid in the Fractional Quantum {{Hall}} Regime},
  author = {Milliken, F. P. and Umbach, C. P. and Webb, R. A.},
  year = {1996},
  date = {1996-01-01},
  journal = {Solid State Commun.},
  volume = {97},
  number = {4},
  pages = {309--313},
  issn = {0038-1098},
  doi = {10.1016/0038-1098(95)00181-6}
}

@article{roddaro_2004_Interedge,
  title = {Interedge {{Strong-to-Weak Scattering Evolution}} at a {{Constriction}} in the {{Fractional Quantum Hall Regime}}},
  author = {Roddaro, Stefano and Pellegrini, Vittorio and Beltram, Fabio and Biasiol, Giorgio and Sorba, Lucia},
  year = {2004},
  date = {2004-07-22},
  journal = {Phys. Rev. Lett.},
  volume = {93},
  number = {4},
  pages = {046801},
  publisher = {American Physical Society},
  doi = {10.1103/PhysRevLett.93.046801}
}

@article{roddaro_2004_Quasiparticle,
  title = {Quasi-Particle Tunneling at a Constriction in a Fractional Quantum {{Hall}} State},
  author = {Roddaro, Stefano and Pellegrini, Vittorio and Beltram, Fabio},
  year = {2004},
  date = {2004-09-01},
  journal = {Solid State Commun.},
  series = {New Advances on Collective Phenomena in One-Dimensional Systems},
  volume = {131},
  number = {9},
  pages = {565--572},
  issn = {0038-1098},
  doi = {10.1016/j.ssc.2004.05.049}
}

@article{picciotto_1997_Direct,
  title = {Direct Observation of a Fractional Charge},
  author = {de-Picciotto, R. and Reznikov, M. and Heiblum, M. and Umansky, V. and Bunin, G. and Mahalu, D.},
  year = {1997},
  date = {1997-09-01},
  journal = {Nature},
  volume = {389},
  number = {6647},
  pages = {162--164},
  issn = {1476-4687},
  doi = {10.1038/38241}
}

@article{miller_2007_Fractional,
  title = {Fractional Quantum {{Hall}} Effect in a Quantum Point Contact at Filling Fraction 5/2},
  author = {Miller, Jeffrey B. and Radu, Iuliana P. and Zumb\"uhl, Dominik M. and Levenson-Falk, Eli M. and Kastner, Marc A. and Marcus, Charles M. and Pfeiffer, Loren N. and West, Ken W.},
  year = {2007},
  date = {2007-08},
  journal = {Nature Physics},
  volume = {3},
  number = {8},
  pages = {561--565},
  publisher = {Nature Publishing Group},
  issn = {1745-2481},
  doi = {10.1038/nphys658}
}

@article{radu_2008_QuasiParticle,
  title = {Quasi-{{Particle Properties}} from {{Tunneling}} in the v = 5/2 {{Fractional Quantum Hall State}}},
  author = {Radu, Iuliana P. and Miller, J. B. and Marcus, C. M. and Kastner, M. A. and Pfeiffer, L. N. and West, K. W.},
  year = {2008},
  date = {2008-05-16},
  journal = {Science},
  volume = {320},
  number = {5878},
  pages = {899--902},
  publisher = {American Association for the Advancement of Science},
  doi = {10.1126/science.1157560}
}

@article{veillon_2024_Observation,
  title = {Observation of the Scaling Dimension of Fractional Quantum {{Hall}} Anyons},
  author = {Veillon, A. and Piquard, C. and Glidic, P. and Sato, Y. and Aassime, A. and Cavanna, A. and Jin, Y. and Gennser, U. and Anthore, A. and Pierre, F.},
  year = {2024},
  date = {2024-08},
  journal = {Nature},
  volume = {632},
  number = {8025},
  pages = {517--521},
  publisher = {Nature Publishing Group},
  issn = {1476-4687},
  doi = {10.1038/s41586-024-07727-z}
}

@article{saminadayar_1997_Observation,
  title = {Observation of the \$\textbackslash mathit\{e\}\textbackslash mathit\{/\}3\$ {{Fractionally Charged Laughlin Quasiparticle}}},
  author = {Saminadayar, L. and Glattli, D. C. and Jin, Y. and Etienne, B.},
  year = {1997},
  date = {1997-09-29},
  journal = {Phys. Rev. Lett.},
  volume = {79},
  number = {13},
  pages = {2526--2529},
  publisher = {American Physical Society},
  doi = {10.1103/PhysRevLett.79.2526}
}

@article{dolev_2008_Observation,
  title = {Observation of a Quarter of an Electron Charge at the {$\nu$} = 5/2 Quantum {{Hall}} State},
  author = {Dolev, M. and Heiblum, M. and Umansky, V. and Stern, Ady and Mahalu, D.},
  year = {2008},
  date = {2008-04},
  journal = {Nature},
  volume = {452},
  number = {7189},
  pages = {829--834},
  publisher = {Nature Publishing Group},
  issn = {1476-4687},
  doi = {10.1038/nature06855},
  issue = {7189}
}

@article{kane_1992_Transmission,
  title = {Transmission through Barriers and Resonant Tunneling in an Interacting One-Dimensional Electron Gas},
  author = {Kane, C. L. and Fisher, Matthew P. A.},
  year = {1992},
  date = {1992-12-15},
  journal = {Phys. Rev. B},
  volume = {46},
  number = {23},
  pages = {15233--15262},
  publisher = {American Physical Society},
  doi = {10.1103/PhysRevB.46.15233}
}

@article{schmid_1983_Diffusion,
  title = {Diffusion and {{Localization}} in a {{Dissipative Quantum System}}},
  author = {Schmid, Albert},
  year = {1983},
  date = {1983-10-24},
  journal = {Phys. Rev. Lett.},
  volume = {51},
  number = {17},
  pages = {1506--1509},
  publisher = {American Physical Society},
  doi = {10.1103/PhysRevLett.51.1506}
}

@article{fisher_1985_Quantum,
  title = {Quantum {{Brownian}} Motion in a Periodic Potential},
  author = {Fisher, Matthew P. A. and Zwerger, Wilhelm},
  year = {1985},
  date = {1985-11-15},
  journal = {Phys. Rev. B},
  volume = {32},
  number = {10},
  pages = {6190--6206},
  publisher = {American Physical Society},
  doi = {10.1103/PhysRevB.32.6190}
}

@article{callan_1992_Phase,
  title = {Phase Diagram of the Dissipative {{Hofstadter}} Model},
  author = {Callan, Curtis G. and Freed, Denise},
  year = {1992},
  date = {1992-05-04},
  journal = {Nucl. Phys. B},
  volume = {374},
  number = {3},
  pages = {543--566},
  issn = {0550-3213},
  doi = {10.1016/0550-3213(92)90400-6}
}

@article{nomura_2001_Strong,
  title = {Strong {{Quasi-Particle Tunneling Study}} in the {{Paired Quantum Hall States}}},
  author = {Nomura, Kentaro and Yoshioka, Daijiro},
  year = {2001},
  date = {2001-12-15},
  journal = {J. Phys. Soc. Jpn.},
  volume = {70},
  number = {12},
  pages = {3632--3635},
  publisher = {The Physical Society of Japan},
  issn = {0031-9015},
  doi = {10.1143/JPSJ.70.3632}
}

@article{fendley_2007_Edge,
  title = {Edge States and Tunneling of Non-{{Abelian}} Quasiparticles in the \$\textbackslash ensuremath\{\textbackslash nu\}=5/2\$ Quantum {{Hall}} State and \$p+ip\$ Superconductors},
  author = {Fendley, Paul and Fisher, Matthew P. A. and Nayak, Chetan},
  year = {2007},
  date = {2007-01-10},
  journal = {Phys. Rev. B},
  volume = {75},
  number = {4},
  pages = {045317},
  publisher = {American Physical Society},
  doi = {10.1103/PhysRevB.75.045317}
}

@article{ito_2012_QuasiParticle,
  title = {Quasi-{{Particle Tunneling}} in {{Anti-Pfaffian Quantum Hall State}}},
  author = {Ito, Toru and Nomura, Kentaro and Shibata, Naokazu},
  year = {2012},
  date = {2012-07-19},
  journal = {J. Phys. Soc. Jpn.},
  volume = {81},
  number = {8},
  pages = {083705},
  publisher = {The Physical Society of Japan},
  issn = {0031-9015},
  doi = {10.1143/JPSJ.81.083705}
}

@article{dec.chamon_1997_Distinct,
  title = {Distinct universal conductances in tunneling to quantum Hall states: The role of contacts},
  author = {de C. Chamon, Claudio and Fradkin, Eduardo},
  year = {1997},
  journal = {Phys. Rev. B},
  volume = {56},
  issue = {4},
  pages = {2012--2025},
  numpages = {0},
  year = {1997},
  month = {Jul},
  publisher = {American Physical Society},
  doi = {10.1103/PhysRevB.56.2012},
  url = {https://link.aps.org/doi/10.1103/PhysRevB.56.2012}
}

@article{fendley_1995_Exact,
  title = {Exact Nonequilibrium Transport through Point Contacts in Quantum Wires and Fractional Quantum {{Hall}} Devices},
  author = {Fendley, P. and Ludwig, A. W. W. and Saleur, H.},
  year = {1995},
  date = {1995-09-15},
  journal = {Phys. Rev. B},
  volume = {52},
  number = {12},
  pages = {8934--8950},
  publisher = {American Physical Society},
  doi = {10.1103/PhysRevB.52.8934}
}

@article{imura_1997_Quantum,
  title = {Quantum Transport in Two-Channel Fractional Quantum {{Hall}} Edges},
  author = {Imura, K. and Nagaosa, N.},
  year = {1997},
  date = {1997-03-15},
  journal = {Phys. Rev. B},
  volume = {55},
  number = {12},
  pages = {7690--7701},
  publisher = {American Physical Society},
  doi = {10.1103/PhysRevB.55.7690}
}

@article{sandler_1998_Andreev,
  title = {Andreev Reflection in the Fractional Quantum {{Hall}} Effect},
  author = {Sandler, Nancy P. and Chamon, Claudio de C. and Fradkin, Eduardo},
  year = {1998},
  date = {1998-05-15},
  journal = {Phys. Rev. B},
  volume = {57},
  number = {19},
  pages = {12324--12332},
  publisher = {American Physical Society},
  doi = {10.1103/PhysRevB.57.12324}
}

@article{hashisaka_2021_Andreev,
  title = {Andreev Reflection of Fractional Quantum {{Hall}} Quasiparticles},
  author = {Hashisaka, M. and Jonckheere, T. and Akiho, T. and Sasaki, S. and Rech, J. and Martin, T. and Muraki, K.},
  year = {2021},
  date = {2021-05-14},
  journal = {Nat. Commun.},
  volume = {12},
  number = {1},
  pages = {2794},
  publisher = {Nature Publishing Group},
  issn = {2041-1723},
  doi = {10.1038/s41467-021-23160-6},
  issue = {1},
}

@article{cohen_2023_Universal,
  title = {Universal Chiral {{Luttinger}} Liquid Behavior in a Graphene Fractional Quantum {{Hall}} Point Contact},
  author = {Cohen, Liam A. and Samuelson, Noah L. and Wang, Taige and Taniguchi, Takashi and Watanabe, Kenji and Zaletel, Michael P. and Young, Andrea F.},
  year = {2023},
  date = {2023-11-03},
  journal = {Science},
  volume = {382},
  number = {6670},
  pages = {542--547},
  publisher = {American Association for the Advancement of Science},
  doi = {10.1126/science.adf9728}
}

@article{ohashi_2022_Andreevlike,
  title = {Andreev-like {{Reflection}} in the {{Pfaffian Fractional Quantum Hall Effect}}},
  author = {Ohashi, Ryoi and Nakai, Ryota and Yokoyama, Takehito and Tanaka, Yukio and Nomura, Kentaro},
  year = {2022},
  date = {2022-12-15},
  journal = {J. Phys. Soc. Jpn.},
  volume = {91},
  number = {12},
  pages = {123703},
  publisher = {The Physical Society of Japan},
  issn = {0031-9015},
  doi = {10.7566/JPSJ.91.123703}
}

@article{mong_2014_Universal,
  title = {Universal {{Topological Quantum Computation}} from a {{Superconductor-Abelian Quantum Hall Heterostructure}}},
  author = {Mong, Roger S. K. and Clarke, David J. and Alicea, Jason and Lindner, Netanel H. and Fendley, Paul and Nayak, Chetan and Oreg, Yuval and Stern, Ady and Berg, Erez and Shtengel, Kirill and Fisher, Matthew P. A.},
  year = {2014},
  date = {2014-03-12},
  journal = {Phys. Rev. X},
  volume = {4},
  number = {1},
  pages = {011036},
  publisher = {American Physical Society},
  doi = {10.1103/PhysRevX.4.011036}
}

@article{belavin_1984_Infinite,
  title = {Infinite Conformal Symmetry in Two-Dimensional Quantum Field Theory},
  author = {Belavin, A.A. and Polyakov, A.M. and Zamolodchikov, A.B.},
  year = {1984},
  date = {1984},
  journal = {Nucl. Phys. B},
  volume = {241},
  number = {2},
  pages = {333--380},
  issn = {0550-3213},
  doi = {10.1016/0550-3213(84)90052-X}
}

@article{kane_1992_Transport,
  title = {Transport in a One-Channel {{Luttinger}} Liquid},
  author = {Kane, C. L. and Fisher, Matthew P. A.},
  year = {1992},
  date = {1992-02-24},
  journal = {Phys. Rev. Lett.},
  volume = {68},
  number = {8},
  pages = {1220--1223},
  publisher = {American Physical Society},
  doi = {10.1103/PhysRevLett.68.1220}
}

@article{furusaki_1993_Resonant,
  title = {Resonant Tunneling in a {{Luttinger}} Liquid},
  author = {Furusaki, Akira and Nagaosa, Naoto},
  year = {1993},
  date = {1993-02-15},
  journal = {Phys. Rev. B},
  volume = {47},
  number = {7},
  pages = {3827--3831},
  publisher = {American Physical Society},
  doi = {10.1103/PhysRevB.47.3827}
}

@article{levin_2007_ParticleHole,
  title = {Particle-{{Hole Symmetry}} and the {{Pfaffian State}}},
  author = {Levin, Michael and Halperin, Bertrand I. and Rosenow, Bernd},
  year = {2007},
  date = {2007-12-06},
  journal = {Phys. Rev. Lett.},
  volume = {99},
  number = {23},
  pages = {236806},
  publisher = {American Physical Society},
  doi = {10.1103/PhysRevLett.99.236806}
}

@article{lee_2007_ParticleHole,
  title = {Particle-{{Hole Symmetry}} and the \$\textbackslash ensuremath\{\textbackslash nu\}=\textbackslash frac\{5\}\{2\}\$ {{Quantum Hall State}}},
  author = {Lee, Sung-Sik and Ryu, Shinsei and Nayak, Chetan and Fisher, Matthew P. A.},
  year = {2007},
  date = {2007-12-06},
  journal = {Phys. Rev. Lett.},
  volume = {99},
  number = {23},
  pages = {236807},
  publisher = {American Physical Society},
  doi = {10.1103/PhysRevLett.99.236807}
}

@article{law_2008_Probing,
  title = {Probing Non-{{Abelian}} Statistics in \$\textbackslash ensuremath\{\textbackslash nu\}=12/5\$ Quantum {{Hall}} State},
  author = {Law, K. T.},
  year = {2008},
  date = {2008-05-12},
  journal = {Phys. Rev. B},
  volume = {77},
  number = {20},
  pages = {205310},
  publisher = {American Physical Society},
  doi = {10.1103/PhysRevB.77.205310}
}

@article{bishara_2008_Quantum,
  title = {Quantum {{Hall}} States at {$\nu$} = 2 k + 2 : {{Analysis}} of the Particle-Hole Conjugates of the General Level- k {{Read-Rezayi}} States},
  shorttitle = {Quantum {{Hall}} States at {$\nu$} = 2 k + 2},
  author = {Bishara, Waheb and Fiete, Gregory A. and Nayak, Chetan},
  year = {2008},
  date = {2008-06-19},
  journal = {Phys. Rev. B},
  volume = {77},
  number = {24},
  pages = {241306},
  issn = {1098-0121, 1550-235X},
  doi = {10.1103/PhysRevB.77.241306}
}

@article{braggio_2012_Quasiparticle,
  title = {Quasiparticle Agglomerates in the {{Read}}--{{Rezayi}} and Anti-{{Read}}--{{Rezayi}} States},
  author = {Braggio, A and Ferraro, D and Magnoli, N},
  year = {2012},
  date = {2012-01-01},
  journal = {Phys. Scr.},
  volume = {2012},
  number = {T151},
  pages = {014052},
  publisher = {IOP Publishing},
  issn = {1402-4896},
  doi = {10.1088/0031-8949/2012/T151/014052}
}

@article{bid_2010_Observation,
  title = {Observation of Neutral Modes in the Fractional Quantum {{Hall}} Regime},
  author = {Bid, Aveek and Ofek, N. and Inoue, H. and Heiblum, M. and Kane, C. L. and Umansky, V. and Mahalu, D.},
  year = {2010},
  date = {2010-07},
  journal = {Nature},
  volume = {466},
  number = {7306},
  pages = {585--590},
  publisher = {Nature Publishing Group},
  issn = {1476-4687},
  doi = {10.1038/nature09277},
  issue = {7306}
}

@article{venkatachalam_2011_Local,
  title = {Local Charge of the {$\nu$} = 5/2 Fractional Quantum {{Hall}} State},
  author = {Venkatachalam, Vivek and Yacoby, Amir and Pfeiffer, Loren and West, Ken},
  year = {2011},
  date = {2011-01},
  journal = {Nature},
  volume = {469},
  number = {7329},
  pages = {185--188},
  publisher = {Nature Publishing Group},
  issn = {1476-4687},
  doi = {10.1038/nature09680}
}

@article{fendley_2009_Boundary,
  title = {Boundary Conformal Field Theory and Tunneling of Edge Quasiparticles in Non-{{Abelian}} Topological States},
  author = {Fendley, Paul and Fisher, Matthew P. A. and Nayak, Chetan},
  year = {2009},
  date = {2009-07-01},
  journal = {Annals of Physics},
  series = {July 2009 {{Special Issue}}},
  volume = {324},
  number = {7},
  pages = {1547--1572},
  issn = {0003-4916},
  doi = {10.1016/j.aop.2009.03.005}
}

@article{caldeira_1981_Influence,
  title = {Influence of {{Dissipation}} on {{Quantum Tunneling}} in {{Macroscopic Systems}}},
  author = {Caldeira, A. O. and Leggett, A. J.},
  year = {1981},
  date = {1981-01-26},
  journal = {Phys. Rev. Lett.},
  volume = {46},
  number = {4},
  pages = {211--214},
  publisher = {American Physical Society},
  doi = {10.1103/PhysRevLett.46.211}
}

@article{mong_2014_Parafermionic,
  title = {Parafermionic Conformal Field Theory on the Lattice},
  author = {Mong, Roger S. K. and Clarke, David J. and Alicea, Jason and Lindner, Netanel H. and Fendley, Paul},
  year = {2014},
  date = {2014-10-27},
  journal = {J. Phys. A: Math. Theor.},
  volume = {47},
  number = {45},
  pages = {452001},
  publisher = {IOP Publishing},
  issn = {1751-8121},
  doi = {10.1088/1751-8113/47/45/452001}
}

@article{fendley_2006_Dynamical,
  title = {Dynamical Disentanglement across a Point Contact in a Non-Abelian Quantum Hall State},
  author = {Fendley, Paul and Fisher, Matthew P. A. and Nayak, Chetan},
  journal = {Phys. Rev. Lett.},
  volume = {97},
  issue = {3},
  pages = {036801},
  numpages = {4},
  year = {2006},
  month = {Jul},
  publisher = {American Physical Society},
  doi = {10.1103/PhysRevLett.97.036801},
  url = {https://link.aps.org/doi/10.1103/PhysRevLett.97.036801}
}

@article{ahmed_2026_Universal,
  author = {Eslam Ahmed and Ryoi Ohashi and Hiroki Isobe and Kentaro Nomura and Yukio Tanaka},
  title = {Universal Transport Theory for Paired Fractional Quantum Hall States in the Quantum Point Contact Geometry},
  journal = {arXiv:2601.08792},
  year = {2026},
  url = {https://arxiv.org/abs/2601.08792},
  primaryClass = {cond-mat.mes-hall}
}

\end{document}